\documentclass{article}

\usepackage{arxiv}

\usepackage[utf8]{inputenc} 
\usepackage[T1]{fontenc}    
\usepackage{hyperref}       
\usepackage{url}            
\usepackage{booktabs}       
\usepackage{amsfonts}       
\usepackage{nicefrac}       
\usepackage{microtype}      
\usepackage{lipsum}		
\usepackage{graphicx}
\usepackage{natbib}
\usepackage{doi}
\usepackage{authblk}
\usepackage{makecell}
\usepackage{wrapfig}
\usepackage{breakcites}
\usepackage{adjustbox}
\usepackage{float}
\usepackage{multirow}
\usepackage{amsmath}
\usepackage{xcolor}
\usepackage{tabularx}
\graphicspath{ {./images/}}
\usepackage{epstopdf}
\usepackage{enumitem}
\usepackage{acronym}

\title{\textbf{CAN3D: Fast 3D Medical Image Segmentation via Compact Context Aggregation}}

\author{
    Wei~Dai$^{1}$\thanks{Principal corresponding author},
    Boyeong~Woo$^{1}$,
    Siyu~Liu$^{1}$,
    Matthew~Marques$^{1}$,
    Craig~B.~Engstrom$^{1}$,
    Peter B.~Greer$^{3}$,\\
    Stuart~Crozier$^{1}$,
    Jason A.~Dowling$^{2}$ ~and~
    Shekhar~S.~Chandra$^{1}$
}

\affil[$^{1}$]{School of Information Technology and Electrical Engineering, The University of Queensland, Australia}
\affil[$^{2}$]{The Australian e-Health Research Centre, CSIRO, Australia}
\affil[$^{3}$]{School of Mathematical and Physical Sciences, The University of Newcastle, Australia}  

\renewcommand{\shorttitle}{W.Dai et al.}

\hypersetup{
pdftitle={CAN3D: Fast 3D Medical Image Segmentation via Compact Context Aggregation},
pdfsubject={},
pdfauthor={Wei~Dai, Boyeong~Woo, Siyu~Liu, Matthew~Marques, Craig~B.~Engstrom, Peter B.~Greer, Stuart~Crozier, Jason A.~Dowling, Shekhar~S.~Chandra},
pdfkeywords={Semantic Segmentation, MRI, Deep Neural Network, CAN3D},
}

\begin{document}
\maketitle

\begin{abstract}
	Direct automatic segmentation of objects from 3D medical imaging, such as magnetic resonance (MR) imaging, is challenging as it often involves accurately identifying a number of individual objects with complex geometries within a large volume under investigation. To address these challenges, most deep learning approaches typically enhance their learning capability by substantially increasing the complexity or the number of trainable parameters within their models. Consequently, these models generally require long inference time on standard workstations operating clinical MR systems and are restricted to high-performance computing hardware due to their large memory requirement. Further, to fit 3D dataset through these large models using limited computer memory, trade-off techniques such as patch-wise training are often used which sacrifice the fine-scale geometric information from input images which could be clinically significant for diagnostic purposes. To address these challenges, we present a compact convolutional neural network with a shallow memory footprint to efficiently reduce the number of model parameters required for state-of-art performance. This is critical for practical employment as most clinical environments only have low-end hardware with limited computing power and memory. The proposed network can maintain data integrity by directly processing large full-size 3D input volumes with no patches required and significantly reduces the computational time required for both training and inference. We also propose a novel loss function with extra shape constraint to improve the accuracy for imbalanced classes in 3D MR images. Compared to other state-of-art approaches (U-Net3D, improved U-Net3D and V-Net), the proposed network reduced the number of parameters up to two orders of magnitude and achieve much faster inference, up to 5 times when predicting with a CPU (instead of GPU). For the open accessed OAI-ZIB knee dataset, the proposed approach achieved dice coefficient accuracy of 0.98 $\pm$ 0.00 and 0.88 $\pm$ 0.04 for femoral bone and tibial cartilage segmentation, respectively, while reducing the mean surface distance error to 0.36 $\pm$ 0.20 mm and 0.29 $\pm$ 0.10 mm when training volume-wise under only 12G VRAM. Our proposed CAN3D demonstrated high accuracy and efficiency on a pelvis 3D MR imaging dataset for prostate cancer consisting of 211 images with expert manual semantic labels (bladder, body, bone, rectum, prostate) that we have also released publicly for scientific use as part of this work\footnote[1]{https://doi.org/10.25919/45t8-p065}

\end{abstract}

\keywords{Semantic Segmentation \and MRI \and Deep Neural Network \and CAN3D}

\section{Introduction}
The segmentation of individual objects and tissues in medical images has become an important process in procedures such as radiation therapy~\citep{Nyholm2014} and clinical assessment~\citep{pollard2008assessment}. Accurate segmentation of objects from \ac{3D} medical images is challenging due to factors such as the complex geometry and relations of various anatomical structures under investigation, variations in tissue signals (contrasts) within and across healthy and disease-effected tissues and highly patient-specific image datasets. 

Deep learning methods, especially the \ac{CNN}, have recently emerged as a popular solution for many \ac{MR} image segmentation problems. Compared to conventional automatic methods, the deep learning-based methods do not require any hand-crafted features and will automatically learn a hierarchy of increasingly sophisticated features directly from data~\citep{Pereira2016}. Inspired by the work of \citet{yu2015CAN}, we present a fully \ac{CNN} based deep learning approach, CAN3D, that utilizes a proposed 3D \ac{CAM} to dramatically reduce memory consumption for sizeable \ac{MR} volumes segmentation. It has a highly compact structure that could perform end-to-end semantic segmentation for full-size \ac{3D} \ac{MR} data with large shape (no patches or slices). By minimizing a model's complexity, the number of parameters required is several orders of magnitude less compared to other networks. Thus, it could achieve state-of-the-art performance using relatively low-end computing hardware and accelerate inference for large 3D input on \ac{CPUs} within just seconds. The main contributions of this work can be summarised as:
\begin{enumerate}
    \item We extended the \ac{CAN} of \citet{yu2015CAN},  previously limited to \ac{2D} images, into a \ac{CAM} for \ac{3D} volumes. The architecture has a highly compact structure with minimal down-sampling operations so that it is designed specifically for semantic segmentation of \ac{3D} \ac{MR} images. It allows for full-size images with large shape to be fed directly into the model during both training and inference. Thus, by preserving fine details from the input, the model can  accurately segment both prominent and imbalanced classes simultaneously in  seconds using \ac{CPUs}. This is a critical need for certain clinical applications. 
  
    \item We proposed a novel hybrid loss termed \ac{DSF}, a combination of \ac{FL}~\citep{lin2017focal} and \ac{DSL}. \ac{DSL} is a computationally efficient form of the original \ac{DSC} loss~\citep{shi2018bayesian} combined with least squares in a single step. This loss has the additional capability to restrain the absolute values of a target during the training without inducing extra computation and could provide high volumetric accuracy while preserving precise surface information globally throughout the contour of the target. It can also accelerate the training process, thus achieving better accuracy than other widely used losses (\ac{DSC}, \ac{MSE}) under the same amount of training time. This proposed loss is proved to be highly efficient, especially for a compact network structure.
    
    \item We have released our 3D image dataset for the pelvis organs and prostate~\citep{pelvisdata2021}, which has expert manual semantic labels (bladder, body, bone, rectum, prostate) of 211 3D MR images, for the performance evaluation of our model. The proposed method achieves state-of-the-art performance with roughly half of the computation time and much fewer parameters (4 to 400 orders of magnitude) than other chosen CNN models. It also performs favourably compared to \citet{ambellan2019automated} on the OAI-ZIB knee dataset presented, with a much simpler pipeline and faster inference. It also dominates both segmentation accuracy and computational time throughout all chosen architectures for the same knee dataset. As a result, our model could be trained for full-size 3D \ac{MR} images directly using GPU with limited video RAM and achieve fast segmentation at full capacity only using the \ac{CPUs}.
    
\end{enumerate}

\section{Background}
\subsection{Conventional Automatic Method}
In the past decades, the automated segmentation of medical images based on gray level feature~\citep{sharma2006computer, ramesh1995thresholding} and texture features~\citep{wang1996comparison, sharma2006computer} has been expanded rapidly in efforts to supersede, with comparable accuracy but significantly increased speed, manual approaches which are time and expertise-intensive and prone to variable (human) intra- and inter-rater reliability. Other automatic segmentation approaches such as model-based and atlas-based approaches have demonstrated significantly faster processing times compared with manual analyses whilst achieving similar measurement accuracy in relation to inter-rater performance and variability as expert observers~\citep{ghose2012survey, ebrahimkhani2020review}.

The use of atlas techniques~\citep{withey2007three} has been used successfully for automated segmentation of anatomical structures such as the prostate~\citep{Klein2008}, brain~\citep{shiee2010topology} and knee~\citep{tamez2011atlas}. \citet{Dowling2015} further showed the application of a multi-atlas approach for accurate automated segmentation of multiple objects (body, bone, prostate, bladder and rectum) from \ac{MR} images of the pelvic region to aid in the generation of pseudo-CT images for \ac{MR}-alone prostate treatment planning. This approach relies on a set of pre-segmented images which serve as exemplars for rigid and non-rigid registrations. The computed transforms are then used to propagate these segmented results from the 'closest matching' atlases, and the final segmentation is obtained by fusing these labels with a fusion method. However, atlas methods are  computationally inefficient requiring a powerful computing cluster to reduce the processing time to several minutes. 

Deformable models have also been successful in the automatic segmentation of \ac{MR} images in many areas, including the (individual) bones~\citep{Schmid2011, Chandra2014a}, muscles~\citep{Engstrom2011}, prostate~\citep{Martin2010} and heart~\citep{Ecabert2008}. Multi-object segmentation based on deformable models has also been developed for both \ac{CT} and \ac{MR} images~\citep{Glocker2012, Chandra2016a}. These models typically involve a triangulated mesh with a shape \textit{prior} learnt from training shapes. The mesh is freely deformed to image features, and the resulting surface is constrained to the correct anatomical shape via the shape model built from the \textit{priors}. They have also been combined with a multi-atlas approach to provide accurate segmentation of the prostate~\citep{Chandra2012c}. However, deformable models are generally susceptible to initialisation error with long computation time, and shape surfaces for model training can be challenging to determine due to anatomical shapes' complex nature. Their performance can also be  limited by a lack of imaging information, such as incomplete or partial fields of view. Solutions to these problems have only recently been proposed~\citep{Schmid2011, Chandra2014a}.

\subsection{Deep Learning Based Automatic Methods}
Inspired by the \ac{FCN}~\citep{Long2015}, encoder-decoder architectures~\citep{noh2015learning, badrinarayanan2017segnet, kendall2015bayesian} such as the 2D U-Net~\citep{Ronneberger2015} and V-Net~\citep{milletari2016v} have been explicitly developed for medical image segmentation. The U-Net has been widely implemented to segment \ac{2D} medical images with losses such as \ac{DSC} for brain tumour~\citep{Dong2017} and \ac{WCE} for cartilage~\citep{Norman2018}. It has also been extended into \ac{3D} volume segmentation for the kidney~\citep{cieck2016} and hand~\citep{kayalibay2017cnn} to utilise the multi-dimensional contextual information fully. \citet{isensee2017brain} further enhanced the U-net3D's performance (Improved U-Net3D) using residual connections~\citep{he2016deep} and deep supervision technique~\citep{kayalibay2017cnn} for 3D brain images. \citet{ambellan2019automated} even incorporate \ac{SSM} with both \ac{2D} and \ac{3D} U-Nets to pass statistical anatomical knowledge during highly pathological knee segmentation. The V-Net~\citep{milletari2016v} is another popular encoder-decoder architecture similar to U-Net. It replaces max-pooling operation with strided convolution for more efficient back-propagation and adds residual blocks for better convergence. The network is trained end-to-end on MRI volumes and can predict segmentation for the whole volume at once. 

Whilst networks such as U-Net and V-Net typically outperform other traditional methods, their complex structures tend to generate numerous parameters which occupy a large portion of \ac{VRAM}, thus, limiting the resources available for input during training and slowing down the computation as more weights need to be updated. Workarounds such as slice-wise~\citep{xue2020multi, wang2019automatic,roth2015improving} or patch-wise~\citep{Dolz2018, dong2020deep, Sun2019} techniques are often applied to save more \ac{VRAM} for the large model by constraining the input data size. However, this could lead to weak feature representation during training because only partial input is utilised through feature extraction. It could also be problematic for fine-scale structures, which may be necessary for justifying these steps during an approval process for clinical adoption and deployment. Furthermore, while these techniques could improve memory efficiency, longer inference time is often required to process all sub-slices or sub-patches sequentially derived from the original input.

Another limitation for encoder-decoder architectures are that the multiple down-sampling layers used to increase the \ac{RF} could also induce losses to the input image's fine-grained information, which are critical for diagnosis. Although the down-sample operations are well suited for classification tasks, the considerable loss of fine-grained resolution could undermine the final performance for dense prediction tasks such as semantic segmentation~\citep{yu2015CAN}. Some recent architectures such as HRNet~\citep{wang2020deep} has been proposed to address this issue by connecting high-to-low resolution convolution streams in parallel and repeatedly exchanging the information across resolutions~\citep{minaee2020image}. However, they still require a considerable number of convolution layers with several image branches, leading to more complex models than baseline encoder-decoder structure. To enlarge the receptive field without suffering information loss or increasing computational cost, dilated convolution has been widely used in real-time segmentation~\citep{yu2015CAN, wang2018understanding, yang2018denseaspp, paszke2016enet}. 

The DeepLab family are among some of the most popular image segmentation methods using dilated convolution. DeepLabv1~\citep{chen2014semantic} and DeepLabv2~\citep{chen2017deeplab} extract features at multiple sampling rates with parallel dilated convolution branches named \ac{ASPP}, thus capturing image features at multiple scales for robust segmentation. Subsequently, DeepLabv3~\citep{chen2017rethinking} combines cascaded and parallel dilated convolution branches together with an improved \ac{ASPP}. The proposed DeepLabv3 framework is then used as an encoder for DeepLabv3+~\citep{chen2018encoder}, which has an encoder-decoder architecture with atrous separable convolution. Dilated convolution has also been utilized for medical images segmentation. \citet{Dolz2018} proposed a U-Net based network with dilated convolution blocks for multi-region segmentation of bladder cancer in \ac{MR} images, and \citet{shi2018bayesian} further implemented it with residual block~\citep{he2016deep} for \ac{3D} heart segmentation. However, none of them focused on the models' efficiency, thus, having their models more complex than the original U-Net. \citet{li2017compactness} is the only work we know so far that trying to reduce the model's complexity for medical image segmentation. However, their work only focused on increasing the residual blocks' efficiency using dilated convolution, and their model is still relatively complex with multiple sets of convolution layers and residual connections, thus limiting their input to small patches even training with a 24G VRAM video card.

\section{Methods}

Data integrity and avoidance of information loss are paramount for medical imaging, as missing critical features could result in poor patient outcomes. Methods that preserve resolution and imaging information would be preferred wherever possible. Thus, the compactness of a neural network, particularly in terms of minimising the number of parameters and down-sampling operations, is crucial in a medical imaging context. Clinical environments seldom have research grade \ac{GPUs} available as vendors typically ship mid-range desktop computers as workstations for instruments such as an \ac{MR} scanner. High-end \ac{GPUs} with large amounts of memory are expensive and require \ac{HPC} environments. Drastically reducing the number of parameters will allow low-end \ac{GPUs}' and even \ac{CPUs}' inference of imaging data to be computed in a few seconds, while the training can still be done on readily available \ac{GPUs} and required only a couple of hours to complete. The reduced capacity of the model could also reduce the likelihood of over-fitting, especially when the training dataset such as \ac{MR} images are commonly limited.~\citep{yu2005integrated} 

\subsection{Network Architecture}

\subsubsection{Progressive Dilated Convolution} 
In medical images, global context indicates how organs and tissues are arranged with respect to its neighbours and is required to obtain precise segmentation of anatomical structures under clinical examination. This global information is typically obtained by increasing the size of receptive fields through the forward pass of a \ac{CNN}, to extract the large-scale features present. The standard approach has been to add down-sampling operations, such as max-pooling, sequentially into the network to project the input onto a low-resolution latent space.  However, this method increases the number of parameters within the network substantially as more layers are needed to regain the spatial information discarded during down-sampling and recover the original image resolution.  Nonetheless, the final performance will decline because the discarded information can never be "fully" restored. Solution such as skip connections~\citep{Ronneberger2015} in U-Net have been proposed to address this problem by propagating gradient information back and forth between the encoder and decoder during training. However, the proposed skip connections still have problems such as forcing aggregation only at the same-scale feature maps which was addressed in \citet{zhou2019unet++}

To cope with the large receptive field required for global context information, the dilated convolution operator, initially developed for wavelet decomposition~\citep{holschneider1990CANFRE}, was adopted for natural image semantic segmentation by \citet{yu2015CAN}, as an alternative to the down-sampling operation in the deep neural network. The key idea is to enlarge the \ac{RF} logically by expanding the kernel with 'holes' (hence also referred to as atrous convolution) instead of physically by decreasing the resolution of the target image. The \ac{3D} atrous convolution operation $*_l$ can be seen as a generalisation of a standard \ac{3D} convolution between \ac{3D} volume $I$ and \ac{3D} filter kernel $k$, with dilation rate specified by a positive integer $d$:
\begin{equation}
    \left(I *_{d} k\right)(p)=\sum_{s+d t=p} I(s) k(t)
\end{equation}

By setting $d = 1$, we obtain the standard \ac{3D} convolution operation. As shown in Figure~\ref{fig:structure}(b), increasing $d$ primarily results in the separation of pixels within a filter by our dilation rate, hence, increasing the receptive field of the filter without an increase in the required number of parameters. The new kernel size after dilation will be:
\begin{equation}
    \hat{u}=u+(u-1)(d-1)
\end{equation}

Where  $u$  and  $\hat{u}$ are the original and dilated kernel size, $d$ is the dilation rate. However, an inherent problem when implementing dilated convolutions is that the aggressively increasing in dilation factors may cause failure to capture local information~\citep{wang2018understanding}. The reason is mainly due to the sparsity of the convolutional kernel after dilation. To address this problem, we followed the strategy proposed in both \citet{yu2015CAN} and \citet{romera2017efficient}, where the dilation rate is slowly and progressively increased along the convolutions within a network, thereby decreasing sparsity in the dilated kernel and allowing more global context to be captured while preserving the resolution of the local region. 

\begin{figure}[!t]
    \centering
    \includegraphics[width=0.8\textwidth]{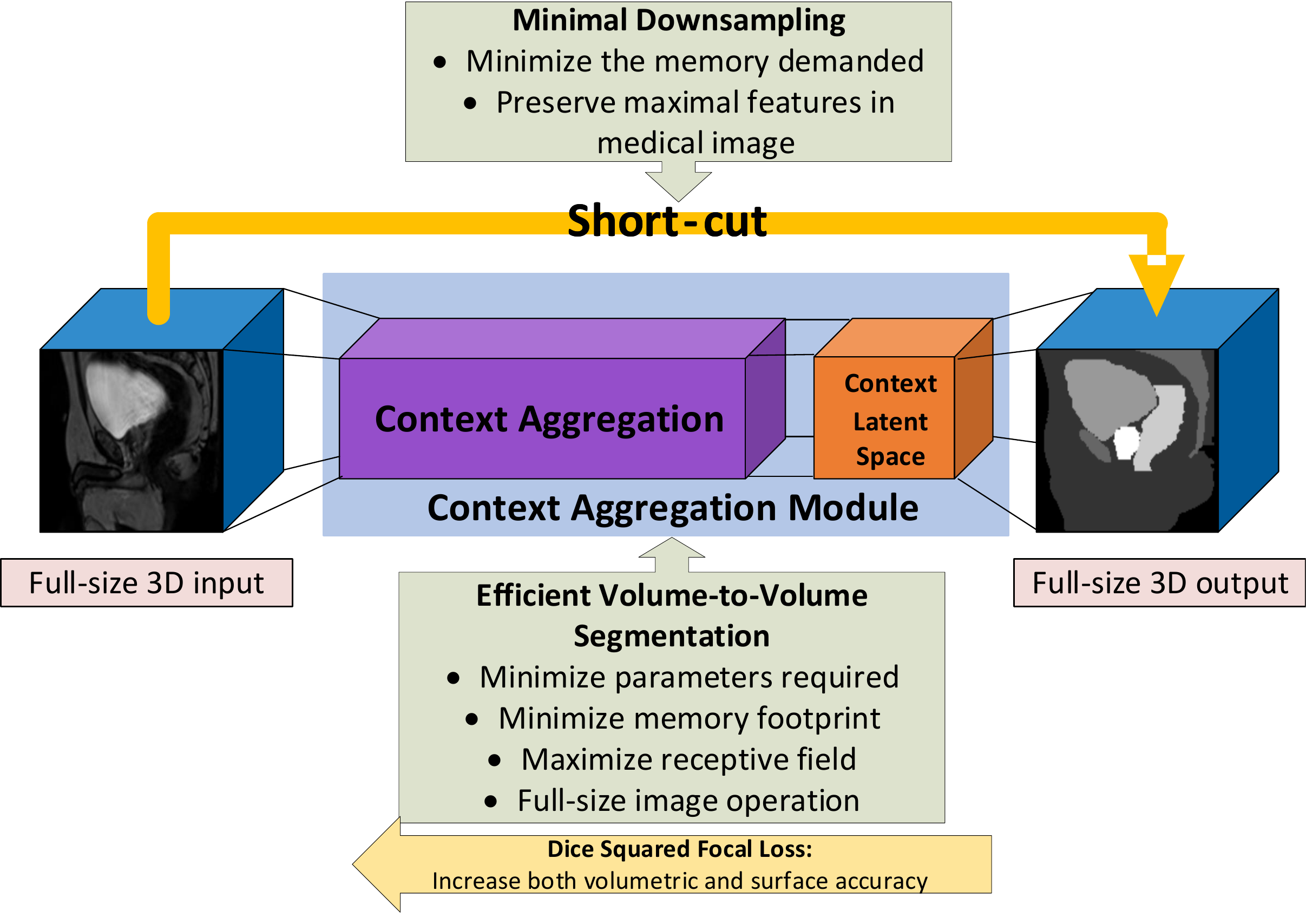}
    \caption{Overview structure of proposed \ac{CAN3D}. }
    \label{fig:overview}
\end{figure}

\subsubsection{Context Aggregation Network(CAN)}
To leverage volumetric medical data with a low number of parameters and allow multi-scale information to be effectively captured for better segmentation, we designed our network based on the \ac{2D} \ac{CAN} proposed in \citet{yu2015CAN}. It utilises a progressively dilated convolution operator to aggregate multi-scale contextual information with no sub-sampling operations. Thus, the network is a context module explicitly design for dense prediction such as semantic segmentation, as it can extract dense feature and perform pixel-wise prediction without degrading the resolution of intermediate representations (no down-sampling required). By systematically connecting multiple convolutional layers with progressively larger dilation rates, the accumulation of variable kernel sizes could expand the overall receptive field exponentially regarding the network depth and allow multi-scale aggregation of contextual information for better localisation. Hence, the full resolution will be considered when predicting each voxel in the final segmentation, regardless of the large image size. 

\subsubsection{CAN3D}
To process volumetric data and exploit high dimensional spatial information, we extend CAN into 3D \ac{CAM} (Figure~\ref{fig:overview}) to achieve efficient volume-to-volume semantic segmentation. To minimize the memory demanded by the model while preserving maximal resolution in the intermediate layers within \ac{CAM}, minimal down-sampling operations with short-cuts (one or two depends on input size) are applied before feeding data into the CAM, so that the layer channels within the \ac{CAM} can be maximized for better feature extraction. The reduction of down-sampling operations can avoid remedial procedures such as up-sampling for reconstruction, so a compact model with a smaller memory footprint could be practically implemented in a clinical situation. Figure~\ref{fig:overview} shows the architecture of the proposed \ac{CAN3D} for volumetric image segmentation. 

\begin{figure*}[!t]
    \centering
    \includegraphics[width=1\textwidth]{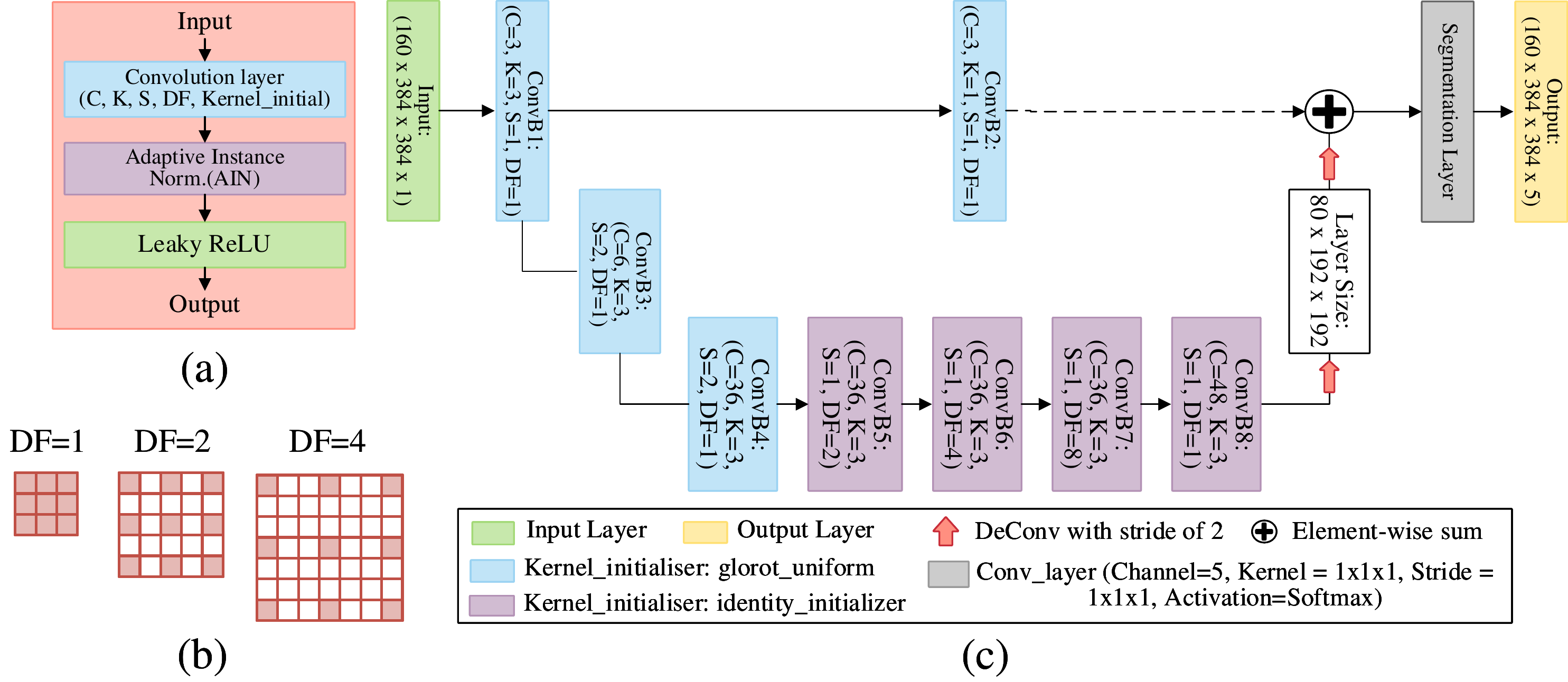}
    \caption{(a) The illustration of single convolution block which consists of an Adaptive Instance Normalization layer, Leaky ReLU activation layer and dilated convolution layer with parameters ($C$, $K$, $S$, $DF$, $Kernel\_Initial$). $C$ is the number of channels, $K$ is the filter size with the shape of $K\times K\times K$, $S$ is the stride size, $DF$ is the dilated factor, and $Kernel\_Initial$ is the kernel initializer used for that convolutional layer. (b) The dilated convolution kernel with the shape of  $3\times3\times3$ under dilation factor of $1$,$2$ and $4$ (c) The proposed compact \ac{CAN3D} network architecture for volumetric image segmentation. The network mainly utilises \ac{3D} context aggregation module (purple blocks) to incorporate features at multiple scales and perform an end-to-end mapping from the image volume to a multi-channel voxel-level dense segmentation. Blue blocks are standard convolution units.}
    \label{fig:structure}
\end{figure*}

The \ac{CAM} which consists of several convolution blocks forms the backbone of our network. As shown in Figure~\ref{fig:structure}(a), each block is composed of an atrous convolution layer with a \ac{LReLU} activation function~\citep{liew2016bounded}, which is a standard activation function $\Phi(x)=\max (\alpha x, x)$, with $\alpha=0.1$ . It avoids the zero-gradient flow problem from standard ReLU and demonstrates improvements in \citet{wang2018classification}. We then employ an \ac{AdaIN} layer between convolution and activation layer, where the \ac{AdaIN} is a linear combination of the identity branch and instance normalisation branch: $\Psi(x)=a x+b \cdot I N(x)$ with weights $a$ and $b$ learned alongside other network parameters. It is similar to \ac{ABN} proposed in \citet{Chen2017} to reduce internal covariance shift and accelerate the convergence. Compared to the standard \ac{IN} layer~\citep{ulyanov2016instance}, two learnable scalar weights are used to balance the strengths of the normalization branch and identity mapping branch, which allows the model to achieve self-adjustment based on operation characteristics. Combing these components, we have:

\begin{equation}
    L_{i}^{S}=\Phi\left(\Psi^{S}\left(b_{i}^{S}+\sum_{j} L_{i,j}^{S-1} *_{r_{S}} K_{i, j}^{S}\right)\right)
\end{equation}

Where $L_i^s$ is the $i^{th}$ feature map of $s^{th}$ block $L$, $b_i^s$ is a scalar bias, and $K_{i,j}^{s}$ is $j^{th}$ channel of convolution kernel, K, for $i^{th}$ feature map.

There are two types of convolutional blocks. The first type is the standard convolutional unit used outside \ac{CAM}. Its dilation rate is set to $1$, and Glorot uniform initializer is used to initialize kernel weights from random distributions. However, this initialization scheme was found ineffective for the second block type, which is a \ac{CAM} unit with the dilation rate set to $r_s=2^{s-1}$. Instead, the identity initializer designed for the recurrent network in \citet{le2015simple} is chosen as an alternative approach to pass the input directly to the output: $k^b(t,a)=1_{t=0}  1_{a=b}$ , where $a$ is the index of input feature map and b is the index of the output map. \citet{yu2015CAN} show that contextual information propagated within the network could be reliably captured by back-propagation during training with the help of an identity initializer, thus, increasing the final segmentation accuracy.

Technical details of the proposed network are shown in the Figure~\ref{fig:structure}(c). Smaller kernels that have been demonstrated to be more efficient for a convolutional network in \citet{tran2015learning} are implemented to capture the minimum low-level features such as edges from a \ac{3D} dataset. All convolution blocks used in the network have a filter size of $3$, except the short-cut layer with $1$. The first convolution block uses a stride of $1$ to increase the channel number of feature maps from $1$ to $8$. For down-sampling, a convolutional block with stride $2$ is applied to halve the feature maps size and increase the channel number to $36$. The sub-sampled feature maps are then fed into the \ac{CAM}, consisting of four convolutional blocks with dilation factors of $2$,$4$,$8$ and $1$, respectively. The last convolution block within \ac{CAM} works as a latent context space to increase the channel number by $48$ for better feature extraction before up-scaling. A $2$-strided deconvolutional block with \ac{AdaIN} and \ac{LReLU} is employed during up-sampling to restore the original resolution. Finally, a segmentation layer with a $1\times1\times1$ convolution and Softmax activation is further applied to generate a multi-channel probability map for final segmentation. 

Overall, our network can accomplish aggregation of global contextual information for high-resolution images with an extremely compact parameterisation. It will minimise the loss of information during training and leads to a smaller memory footprint for volume-to-volume segmentation which could be practically implemented in a clinical situation. Its minimal sub-sampling operation also prevents the possibility of over down-sampling; thus, make it possible to handle various input shapes with no customization required.

\subsection{Loss Metric}
High class imbalance is another common challenge for medical image segmentation. Using the pelvis dataset in this experiment as an example, the number of voxels in the background are approximately $270$ times greater than the prostate. Thus, we propose a novel hybrid loss named Dice Squared Focal loss (\ac{DSF}), a combination of our proposed Dice Square Loss (\ac{DSL}) and an Focal Loss (\ac{FL})~\citep{lin2017focal} proposed for dense object detection.
\begin{equation}
    L_{D S F}=L_{D S L}+L_{F L}
\end{equation}

\subsubsection{Dice Similarity Coefficient Loss}
The \ac{DSC}~\citep{cieck2016} gauges the similarity of segmented images and manual delineated images based on their overlap measurement. Its loss is usually defined as:
\begin{equation}
    L_{D S C} = 1 - S_{D S C}=1 - \frac{\sum_{i \in I} P_{i} Q_{i}}{\sum_{i \in I} P_{i}^{2}+\sum_{i \in I} Q_{i}^{2}}
\end{equation}
Where $P$ is the ground truth set and $Q$ is the estimated segmentation set. $P$ and $Q$ have the same shape with a total number of voxels, $I$.  Even though \ac{DSC} loss is an excellent regional loss to increase the overlap between the target and the result, it is not specific to the absolute values of the sets. The \ac{MSE}, on the other hand, is specific to the absolute values of the target without taking the regional errors into account explicitly. Therefore, a better version of \ac{DSC} loss is needed to keep both regional and absolute values errors from diverging during the optimization, thus, stabilizing the training.

\subsubsection{Dice Squared Focal Loss}
Inspired by the loss form of DSC, we introduce \ac{DSL}, which is an efficient form of DSC loss with \ac{MSE} built inside. It inherits \ac{DSC}'s superiority by increasing the overlap between two segmented volumes while restraining the absolute values' difference by calculating an additional \ac{MSE} term for free. Inspired by the similarity definitions of \citet{cha2007comprehensive}, an alternative expression for \ac{DSC} loss can be defined as:
\begin{equation}
    L_{D S C}=1-S_{D S C}=\frac{\sum_{i \in I}\left(P_{i}-Q_{i}\right)^{2}}{\sum_{i \in I} P_{i}^{2}+\sum_{i \in I} Q_{i}^{2}}
\label{eq:dsc}
\end{equation}

From the equation, we can find the $L_{DSC}$ already includes the square difference error during its calculation, which means an extra \ac{MSE} term can be calculated with only one additional floating-point operation (a division by overall voxel number). A weighted multi-channel \ac{DSL} can be calculated as:

\begin{equation}
    L_{D S L}=\frac{\sum_{k \in K}\left(w_{k} *\left(\frac{\Psi}{\sum_{i \in I} P_{i, k}^{2}+\sum_{i \in I} Q_{i, k}^{2}}+\frac{\Psi}{I}\right)\right)}{\sum_{k \in K} w_{k}}
\end{equation}
where $\Psi=\sum_{i \in I}\left(P_{i, k}-Q_{i, k}\right)^{2}$ is the square difference sum between multi-channel output ($P$), and one-hot-encoded ground truth ($Q$). $K$ equals to the number of classes, $w$ is weighting factor for each class and I is the total number of voxels. Even though \ac{DSL} can smooth segmentation results globally, it is insensitive to local uncertainties of the surface of segmented volume. Therefore, we further introduce \ac{DSF} by summing our \ac{DSL} with a weighted \ac{FL}~\citep{lin2017focal} (factor of 10). The \ac{FL} is shown as:
 \begin{equation}
     L_{F L}=\sum_{k \in K}-\alpha_{k}\left(1-p_{k}\right)^{\gamma} \log \left(p_{k}\right)
 \end{equation}
where $\alpha$ is a weighting factor to adaptively balance the variant of this loss and p is the estimated probability for each class k. In general, $(1-p_{k} )^\gamma$ is a scaling factor to down-weight the relative loss for well-classified examples and make the model focus more on the misclassified voxels. Focal loss tends to preserve intricate boundary details and is mainly used to complement with \ac{DSL} for better accuracy on the contour of segmentation.

\section{Experiments}

\subsection{Data}

Our proposed model was evaluated on two MRI datasets with large image shapes, a public knee dataset from \ac{OAI}-\ac{ZIB}~\citep{seim2010model} and a private pelvis dataset~\citep{dowling2015automatic} obtained from 38 prostate cancer patients to demonstrate its capability to segment large-scale multi-dimensional images. Both datasets' manual segmentation are used as ground truth label for both model training and performance evaluation.

The \ac{OAI}-\ac{ZIB} knee dataset with exceptionally large image shape were used to push the proposed model's performance up to an extreme situation under restricted \ac{VRAM}. It consists of MR images from the public \ac{OAI} database, whose manual segmentation for 507 baseline images were carried out by experienced examiners at \ac{ZIB}~\citep{ambellan2019automated}. All images were acquired using a Siemens 3T Trio with an image resolution of $0.36\times0.36\times0.7$ mm and a shape of $160\times384\times394$. The manual segmentation consists of 5 difference classes: background (1), \ac{FB} (2), \ac{TB} (3), \ac{FC} (4) and \ac{TC} (5).

The proposed model was also evaluated with a private pelvis dataset which has been released by us for this paper~\citep{pelvisdata2021}. The dataset consists of 211 3D MR images of thirty-eight patients aged between 58-78 years, who had been diagnosed with tumours. \ac{MR} image sequences for each patient were acquired with a Siemens Skyra 3T scanner, and all patients were positioned by two radiation therapists. The planning MR was a \ac{3D}, T2 weighted 1.6 mm isotropic \ac{SPACE} sequence with large field-of-view to cover the entire pelvis with bladder, and the prostate delineation sequence was a \ac{2D} axial T2 weighted sequence with small field-of-view approximately $200\times200$ $mm^2$. All classes except prostate were manually segmented for all patients independently from the large FOV T2 scan. The prostate was manually segmented from the small FOV T2 scan and then propagated to the large FOV T2 MR image. All manual segmentation was conducted by two experienced radiation oncologists and an experienced research radiation therapist. The final segmentation used for training and validation were obtained by combining the segmentation from different observers using majority voting. The image size of the input is $256\times256\times128$, and the average voxel size is approximately $1.68\times1.68\times1.56$ mm. Overall, manual segmentation consist of six different classes which are available for each patient: background (1), bladder (2), body (3), bone (4), rectum (5) and prostate (6).

With \ac{MR} image intensity values being non-standardized, the data acquired needs to be normalized to match the range of values and avoid the network's initial biases. Thus, both datasets were normalized to have a zero mean and unit variance. Moreover, N4 bias field correction~\citep{tustison2010n4itk} was also implemented to correct low-frequency intensity non-uniformity. 
\subsection{Evaluation}
The similarity between predicted segmentation and ground truth is assessed by employing \ac{DSC} (Eq.\ref{eq:dsc}) to compare both volumes based on their overlap. However, volume-based metrics generally lack sensitivity to errors on the segmentation's contour if the segmented volume is relatively large, which is particularly crucial to clinical applications such as radiation treatment planning. Hence, both \ac{MSD} and \ac{HD}~\citep{Schmid2011}, which are sensitive to variance on the anatomical boundaries, are also utilized to quantify the performance of segmentation. Both metrics are designed based on the surface distance shown as:

\begin{equation}
    d(p, Q)=\min _{q \in Q}|| p-q \|_{2}
\end{equation}

Where $p$ and $q$ are points on surfaces of predicted volume, $P$, and ground truth volume, $Q$, respectively. $|| .\|_{2}$ calculates the Euclidian distance between two points. Surface distance calculates the minimum distance between each point from one surface to another. By averaging surface distance for all points between two surfaces, mean surface distance is obtained as:

\begin{equation}
    MSD(P, Q)=\frac{1}{n_{p}+n_{q}}\left(\sum_{p \in Q} d(p, Q)+\sum_{q \in Q} d(q, P)\right)
\end{equation}

Where $n_p$ and $n_q$ are the total number of points on surface $P$ and $Q$. The Hausdorff distance, on the other hand, finds the longest surface distance between two surfaces. The symmetric Hausdorff distance is defined as:

\begin{equation}
    HD(P, Q)=\max \left(\max _{p \in P} d(p, Q), \max _{q \in Q} d(q, P)\right)
\end{equation}

\subsection{Implementation}

Three-fold cross-validation studies were conducted for both datasets. For the \ac{OAI}-\ac{ZIB} knee dataset, the overall 1012 images were sorted numerically and split into three sets with a portion of $(338/338/336)$. All results recorded in this paper are averaged three-fold results. For the pelvis dataset, the overall $211$ MR images from $38$ cases were decomposed into three subsets by random choice $(76/81/54)$. Offline augmentation strategies such as elastic deformation, affine wrapping, uniaxial rotation and shifting were applied to enlarge the training set and avoid the potential of over-fitting. 

For network training, the proposed \ac{DSF} was used as the loss function for our model, while Adam optimization algorithm with standard beta values of $0.9$ and $0.999$ was applied to minimize this loss. The polynomial decay strategy was utilised to exponentially decrease the learning rate with a power of two after each epoch until reaching around zero at the end of training. An initial learning rate of $0.001$ was specifically used for CAN3D. During the training, only one image with the full resolution was employed for each batch due to the specified \ac{VRAM} limitation. Experiments were implemented based on TensorFlow 2.1.0 version, and the training of all models were conducted on an NVIDIA Volta V100 GPU with 32GBs of memory. To test segmentation performance of models under limited resources, the overall GPU \ac{VRAM} was further capped to 12GBs to simulate low-end computing equipment. This approach has been tested to provide similar results with faster training time compared to Nvidia Titan V, which has 12GBs RAM natively. Once the models were trained, they were tested using both mentioned GPU and a separate CPU (Intel Xeon Gold 613) for performance evaluation under different hardware conditions.

Multiple deep learning architectures such as the U-Net3D~\citep{cieck2016}, improved U-Net3D~\citep{isensee2017brain} and V-Net~\citep{milletari2016v} were chosen as the state-of-art 3D models for comparison. Since none of these networks' original structure could fit the chosen datasets with full resolution under 12G \ac{VRAM}, all networks had their filter numbers adjusted within each convolution layer to provide the best performance possible under the specified \ac{VRAM} limitation. All models were also trained from scratch under the same epoch numbers with their original hyper-parameters untouched, except that the same optimizer and learning rate scheduler were used as training control.

Three experiments were conducted to evaluate the proposed loss function and model with two different dataset under different hardware scenarios:
\begin{itemize}
    \item \textbf{Loss Function:} To evaluate the proposed loss function's performance, the two most popular losses for image segmentation, weighted \ac{CE} and weighted \ac{DSC}, were evaluated along with the proposed \ac{DSL} and \ac{DSF} using the CAN3D. All losses were trained until converging when the improvement of accuracy is less than 0.005 within 5 epochs.To demonstrate the performance of \ac{DSF} on different network structures, \ac{DSF} was also tested using other chosen models, and the results are evaluated with the original losses of these models.
    \item \textbf{OAI-ZIB Knee Dataset:} All models were tested with their proposed losses using the OAI-ZIB knee dataset under 12G \ac{VRAM}, to demonstrate the efficiency of CAN3D for large input under the low-end computing environment.  Moreover, to demonstrate the full potential of CAN3D, the proposed model was also trained under full hardware capacity (32 \ac{VRAM}) with the stride of \ac{CAM}'s ConvB4 block (shown in figure \ref{fig:structure}c) reduced to one so the layer shape could be increased throughout the \ac{CAM}. The results generated from this full capacity test were then evaluated with the results of \citet{ambellan2019automated}. Since \citet{ambellan2019automated} used shape model for post-processing, more straightforward techniques involving the largest connected component extraction and holes filling were applied to post-process CAN3D's output for a fair comparison.
    \item \textbf{Pelvis Dataset:} When fitting the knee dataset during the second experiment,  all models' parameters except CAN3D were truncated dramatically due to the large shape of input, and some models failed to work entirely. Therefore, a pelvis dataset with a much smaller image shape was also tested to compare the efficiency of CAN3D with other models when all models could generate some primary results. A similar CAN3D structure as the full capacity test in the knee dataset experiment was used with its ConvB1(shown in figure~\ref{fig:structure}c) block's filter number increased to eight. For other models, the filter number of convolutional layers were increased accordingly to adapt to the smaller dataset. Moreover, pelvis dataset results from two traditional automatic methods, a multi-atlas approach reported by \citet{Dowling2015} and a multi-object weighted deformable model proposed by \citet{Chandra2016a}, and were also provided for further evaluation.
\end{itemize}
\section{Results}
\subsection{Experiment I: Loss function}

\begin{table}[!t]
\centering
\setlength{\tabcolsep}{3.5pt}
\scalebox{0.85}{
\begin{tabular}{ccclc}
\hline
Metric & DSC       & HD (mm)   & MSD (mm)  & Time (s/image) \\ \hline
CE     & 0.81±0.06$^*$ & 9.80±4.31$^*$ & 2.46±0.97$^*$ & 0.835          \\
DSC    & 0.83±0.06$^*$ & 16.9±24.4$^*$ & 1.99±0.78$^*$ & 0.834          \\
DSL & \multicolumn{1}{l}{0.85±0.04$^*$} & \multicolumn{1}{l}{13.2±18.4$^*$} & 1.91±0.72$^*$          & \textbf{0.789} \\
DSF & \textbf{0.86±0.05}             & \textbf{8.87±7.51}            & \textbf{1.76±0.73} & 0.817          \\ \hline
\end{tabular}
}
\caption{Performance of different loss functions trained using the proposed model. All losses are trained until converging with a delta of 0.005. Time represents the training time and all losses are weighted with same class weights. * represent metrics with p-value $<$0.01 (two-sided paired wilxocon signed-rank test) comparing to the DSF, Bold represent best score within each metric.}
\label{Table:loss}

\vspace{0.1cm}

\setlength{\tabcolsep}{3.5pt}
\scalebox{0.8}{
\begin{tabular}{cc|cc|ccl}
\cline{1-6}
Region &
  \begin{tabular}[c]{@{}c@{}}Mean \\ Metric\end{tabular} &
  \begin{tabular}[c]{@{}c@{}}Cicek et.al.\\ (WCE)\end{tabular} &
  \begin{tabular}[c]{@{}c@{}}Cicek et.al.\\ (DSF)\end{tabular} &
  \begin{tabular}[c]{@{}c@{}}Isensee et.al.\\ (WDSC)\end{tabular} &
  \begin{tabular}[c]{@{}c@{}}Isensee et.al.\\ (DSF)\end{tabular} &
   \\ \cline{1-6}
         & DSC & 0.948±0.044$^*$                        & \textbf{0.978±0.011} & 0.980±0.011$^*$                        & \textbf{0.982±0.009} &  \\
Body     & HD  & 46.78±25.98$^*$                        & \textbf{43.29±27.02} & 36.01±12.12$^*$                        & \textbf{25.13±10.89} &  \\
         & MSD & 4.438±4.267$^*$                        & \textbf{1.154±0.574} & 1.079±0.590$^*$                        & \textbf{0.918±0.458} &  \\ \cline{1-6}
         & DSC & 0.855±0.173                        & \textbf{0.867±0.170} & 0.919±0.089$^*$                        & \textbf{0.917±0.093} &  \\
Bladder  & HD  & 32.18±39.26$^*$                        & \textbf{23.35±35.82} & 28.08±36.07                        & \textbf{21.06±31.01} &  \\
         & MSD & \textbf{5.276±9.552$^*$} & 6.584±18.63                        & \textbf{2.301±3.540$^*$} & 3.163±5.561                        &  \\ \cline{1-6}
         & DSC & 0.907±0.021                        & 0.904±0.025                        & 0.913±0.020$^*$                         & \textbf{0.914±0.020} &  \\
Bone     & HD  & 36.23±22.81                        & \textbf{33.25±19.76} & 41.73±28.95$^*$                        & \textbf{22.03±12.63} &  \\
         & MSD & \textbf{1.653±0.427$^*$} & 1.694±0.542                        & 1.533±0.391                        & \textbf{1.492±0.372} &  \\ \cline{1-6}
         & DSC & 0.779±0.101$^*$                        & \textbf{0.821±0.072} & \textbf{0.866±0.050}  & 0.850±0.061                        &  \\
Rectum   & HD  & 27.12±26.61$^*$                        & \textbf{23.79±26.72} & \textbf{19.18±25.00} & 19.51±21.28                        &  \\
         & MSD & 3.221±2.837$^*$                        & \textbf{2.377±1.602} & \textbf{1.739±1.052} & 1.998±1.191                        &  \\ \cline{1-6}
         & DSC & 0.735±0.158$^*$                        & \textbf{0.783±0.138} & \textbf{0.839±0.069$^*$} & 0.819±0.089                        &  \\
Prostate & HD  & 31.99±54.62$^*$                        & \textbf{13.37±23.68} & \textbf{8.890±8.867$^*$} & 12.83±17.85                        &  \\
         & MSD & 19.76±41.76$^*$                        & \textbf{4.776±15.41} & \textbf{1.920±0.771$^*$} & 2.314±1.723                        & \\ \cline{1-6}
\end{tabular}}
\caption{Evaluation of the proposed loss function, \ac{DSF}, under U-Net3D~\citep{cieck2016} and improved U-Net3D~\citep{isensee2017brain}. The performance of \ac{DSF} is compared with the original losses of those model for all classes using three-fold cross-validation under the same training time. * represent metrics with p-value $<$0.05 (two-sided paired wilxocon signed-rank test) comparing to the DSF, Bold represent best score within each model.}
\label{Table:loss_other_model}
\end{table}

To illustrate the performance of the different loss functions, particularly for imbalanced classes, results on the various accuracy measurements for the CAN3D segmented prostate are presented in Table 1. In general, both of our \ac{DSL} related losses outperform other losses, especially for \ac{DSC} and \ac{MSD}. When comparing to \ac{DSC} specifically, the \ac{DSL} improves the final segmentation accuracy with even faster training time for each image. While \ac{DSL} and \ac{DSF} achieve close results on \ac{DSC}, the \ac{DSF}, which has an extra \ac{FL} component, can achieve superior \ac{HD}. Although the extra \ac{FL} component introduces some burden to computation speed, the per-image training speed for \ac{DSF} is still faster than \ac{CE} and \ac{DSC}. Moreover, \ac{DSL}'s results are also found to be significantly different from \ac{DSF} with p-values (Wilxocon rank test) for all metrics below 0.01. 

The proposed \ac{DSF} was also tested with the U-Net3D~\citep{cieck2016} and the improved U-Net3D~\citep{isensee2017brain}, and Table \ref{Table:loss_other_model} compares the performance of \ac{DSF} with the original losses of those models. For simpler networks such as the original U-Net3D, dominant performance is observed for the \ac{DSF} compared to the network's original loss, weighted \ac{CE}. For the improved U-Net3D whose original loss is weighted \ac{DSC}, the \ac{DSF} also improves the accuracy of larger components such as body, bladder and bone with a significant difference, especially HD. While the \ac{DSF}'s performance on imbalanced classes such as prostate and rectum is slightly worse than the original weighted \ac{DSC}, the difference in term of rectum's results are not significant with all p-values above 0.05.

\subsection{Experiment II: OAI-ZIB dataset}

\begin{figure*}[!t]
    \centering
    \includegraphics[width=0.8\textwidth]{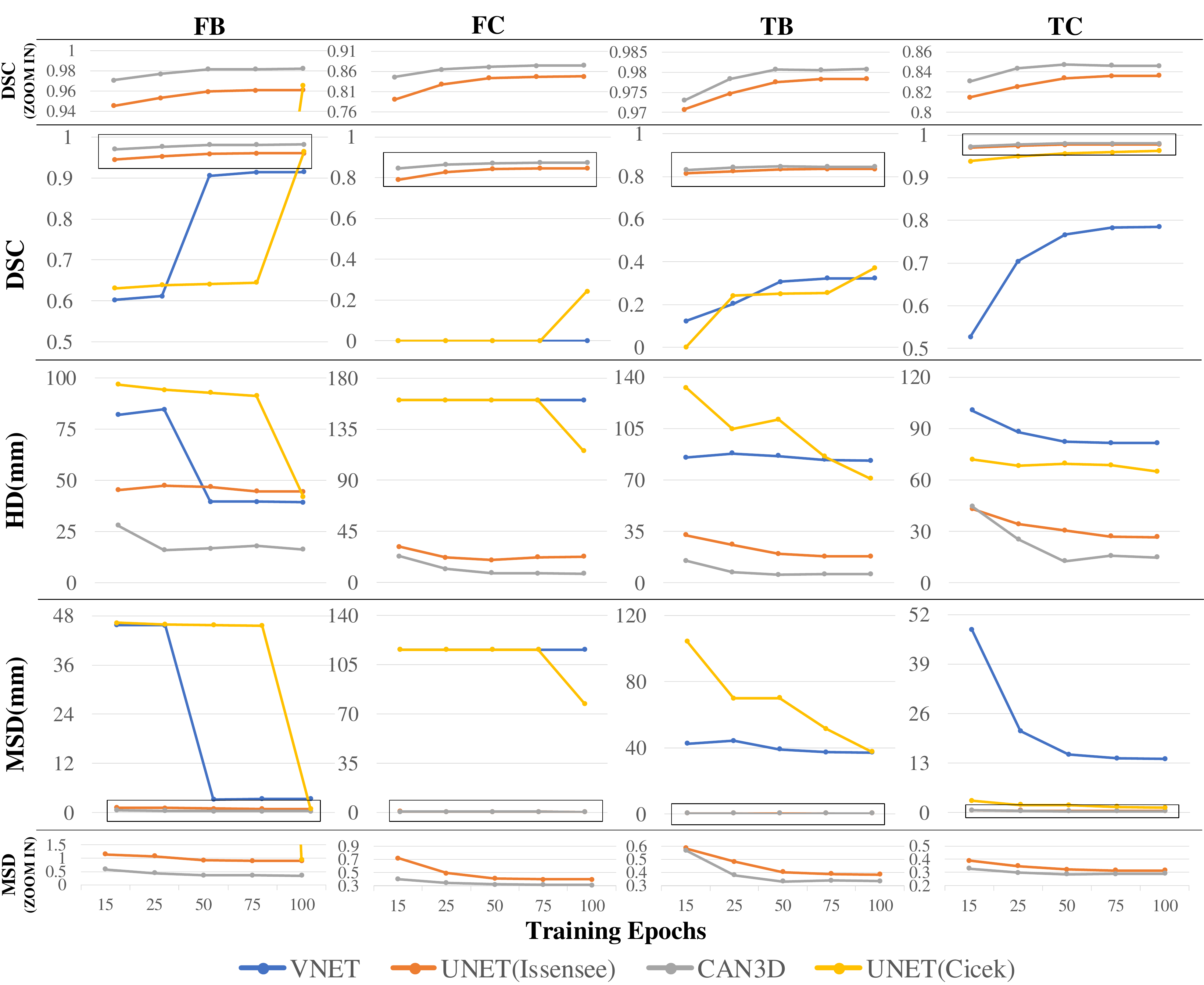}
    \caption{OAI-ZIB Knee dataset performance of VNET~\citep{milletari2016v}, UNET3D~\citep{cieck2016}, improved UNET3D~\citep{isensee2017brain} and CAN3D trained under 15, 25, 50, 75 and 100 epochs. Metrics such as Dice similarity coefficient (DSC), Hausdorff Distance(HD) and mean surface distance (MSD) for each class are plotted separately with single data point representing the averaged three-fold cross-validation results.}
    \label{fig:zib_large}
\end{figure*}

Figure~\ref{fig:zib_large} outlines the models performance under various training epochs for the OAI-ZIB knee dataset using 12G \ac{VRAM}. It can be noted that half of the chosen models barely worked for this dataset as their performances were effected by the truncation of parameters within their model. For instance, the V-Net produces no results for prominent classes such as \ac{FB} and \ac{TB} until reaching 50 epochs and fails completely for the imbalanced classes such as \ac{FC} and \ac{TC}. The U-Net3D is not functional for most classes until trained for 100 iterations. On the other hand, both the improved U-Net3D and CAN3D accomplish great results through all epochs, and a zoomed-in plot for the outlined region is shown at the top and bottom section of figure~\ref{fig:zib_large} for a better comparison.  From the zoomed-in plot, CAN3D can be found surpassing improved U-Net3D in every category throughout all epochs, and its peak performance exceeds the rest of the models by a huge step for all metrics.

\begin{table}[!t]
    \centering
    \setlength{\tabcolsep}{3pt}    
    \scalebox{0.85}{
\begin{tabular}{cccccc}
\hline
{\color[HTML]{000000} Region} &
  {\color[HTML]{000000} Mean Metric} &
  {\color[HTML]{000000} \begin{tabular}[c]{@{}c@{}}Ambellan\\ et al.\end{tabular}} &
  {\color[HTML]{000000} \begin{tabular}[c]{@{}c@{}}CAN3D\\ 12G+post\end{tabular}} &
  {\color[HTML]{000000} \begin{tabular}[c]{@{}c@{}}CAN3D\\ 32G\end{tabular}} &
  {\color[HTML]{000000} \begin{tabular}[c]{@{}c@{}}CAN3D\\ 32G+post\end{tabular}} \\ \hline
{\color[HTML]{000000} } &
  {\color[HTML]{000000} DSC} &
  {\color[HTML]{000000} 0.99±0.30} &
  {\color[HTML]{000000} 0.98±0.00} &
  {\color[HTML]{000000} \textbf{0.99±0.00}} &
  {\color[HTML]{000000} \textbf{0.99±0.00}} \\
\multirow{-2}{*}{{\color[HTML]{000000} TC}} &
  {\color[HTML]{000000} HD (mm)} &
  {\color[HTML]{000000} \textbf{2.93±1.24}} &
  {\color[HTML]{000000} 4.90±2.11} &
  {\color[HTML]{000000} 16.3±16.4} &
  {\color[HTML]{000000} 4.36±2.04} \\ \hline
{\color[HTML]{000000} } &
  {\color[HTML]{000000} DSC} &
  {\color[HTML]{000000} \textbf{0.90±3.60}} &
  {\color[HTML]{000000} 0.87±0.03} &
  {\color[HTML]{000000} 0.89±0.02} &
  {\color[HTML]{000000} 0.89±0.02} \\
\multirow{-2}{*}{{\color[HTML]{000000} FC}} &
  {\color[HTML]{000000} HD (mm)} &
  {\color[HTML]{000000} \textbf{5.35±2.50}} &
  {\color[HTML]{000000} 5.89±3.37} &
  {\color[HTML]{000000} 8.66±9.88} &
  {\color[HTML]{000000} 5.62±3.30} \\ \hline
{\color[HTML]{000000} } &
  {\color[HTML]{000000} DSC} &
  {\color[HTML]{000000} \textbf{0.99±0.33}} &
  {\color[HTML]{000000} 0.98±0.00} &
  {\color[HTML]{000000} 0.98±0.00} &
  {\color[HTML]{000000} 0.98±0.00} \\
\multirow{-2}{*}{{\color[HTML]{000000} TB}} &
  {\color[HTML]{000000} HD (mm)} &
  {\color[HTML]{000000} \textbf{3.16±2.03}} &
  {\color[HTML]{000000} 4.31±2.54} &
  {\color[HTML]{000000} 12.9±17.2} &
  {\color[HTML]{000000} 3.91±2.17} \\ \hline
{\color[HTML]{000000} } &
  {\color[HTML]{000000} DSC} &
  {\color[HTML]{000000} \textbf{0.86±4.54}} &
  {\color[HTML]{000000} 0.85±0.04} &
  {\color[HTML]{000000} 0.85±0.04} &
  {\color[HTML]{000000} 0.85±0.04} \\
\multirow{-2}{*}{{\color[HTML]{000000} TC}} &
  {\color[HTML]{000000} HD (mm)} &
  {\color[HTML]{000000} 6.35±4.36} &
  {\color[HTML]{000000} 5.04±2.55} &
  {\color[HTML]{000000} 5.31±2.75} &
  {\color[HTML]{000000} \textbf{5.04±1.99}} \\ \hline
\multicolumn{2}{c}{{\color[HTML]{000000} Time (s/step)}} &
  {\color[HTML]{000000} \textgreater{}3600} &
  {\color[HTML]{000000} \textbf{10.6}} &
  {\color[HTML]{000000} 11} &
  {\color[HTML]{000000} 15.2} \\ \hline
\end{tabular}
    }
    \caption{OAI-ZIB performance under 32G VRAM before and after post-processing. Time is the inference time for each image using CPU. Bold represents the best score}
    \label{Table:knee_32g}
\end{table}

Due to the various training speeds listed in Table~\ref{Table:zib_result}, even under the same training epochs, the discrepancy among the overall training times spent by different models would cause an unfair comparison when evaluating the models' efficiency. Thus, a further experiment was conducted to run all models under identical training time, and Table~\ref{Table:zib_result} presents the quantitative results when all models are trained for eight hours. It demonstrates that the proposed method obtains superior results for every measurement with a significant difference, especially for the imbalanced classes such as \ac{FC} and \ac{TC}. Since both the U-Net3D and V-Net almost fail to work with this dataset, the performance of CAN3D will only be compared with improved U-Net3D here. Our model achieves a \ac{DSC} of (0.981, 0.847) for both tibial components, which are only slightly better than the improved U-Net3D (0.978, 0.834). However, it also reaches a \ac{DSC} of (0.982, 0.871) for both femoral components which are (0.023, 0.027) higher than the improved U-Net3D. The CAN3D's edge in large volume segmentation is amplified even more when considering the metrics related to the surface distance, as it achieves much better \ac{HD} and \ac{MSD} than improved U-Net3D for all classes, especially for imbalanced classes such as \ac{TC} and \ac{FC}. Moreover, the CAN3D also shows its efficiency by dominating the CPU inference time (1.5 to 2 times faster) despite having more parameters than other models.

The results of CAN3D without memory limitation are summarized and compared with \citet{ambellan2019automated} in Table~\ref{Table:knee_32g}. A large  improvement for \ac{HD} is found for CAN3D after some simple post-processing techniques,  and the post-processed results achieve similar \ac{DSC} as Ambellan at al. Although CAN3D's surface distances are slightly worse than Ambellan at al., it has much simpler pipeline with 200 times faster inference using CPU. Besides, \ac{CAN3D} performs exceptionally well for the highly imbalanced class such as TC.

\begin{table*}[!t]
\centering
\scalebox{0.9}{
\begin{tabular}{cccccc}
\hline
{\color[HTML]{000000} Region} &
  {\color[HTML]{000000} Mean Metric} &
  {\color[HTML]{000000} UNET3D (Cicek)} &
  {\color[HTML]{000000} UNET3D (Issensee)} &
  {\color[HTML]{000000} VNET} &
  {\color[HTML]{000000} CAN3D} \\ \hline
{\color[HTML]{000000} } &
  {\color[HTML]{000000} DSC} &
  {\color[HTML]{000000} 0.641±0.009*} &
  {\color[HTML]{000000} 0.959±0.011*} &
  {\color[HTML]{000000} 0.906±0.027*} &
  {\color[HTML]{000000} \textbf{0.982±0.005}} \\
{\color[HTML]{000000} FB} &
  {\color[HTML]{000000} HD (mm)} &
  {\color[HTML]{000000} 92.95±11.38*} &
  {\color[HTML]{000000} 46.81±14.66*} &
  {\color[HTML]{000000} 39.68±12.19*} &
  {\color[HTML]{000000} \textbf{16.75±15.89}} \\
{\color[HTML]{000000} } &
  {\color[HTML]{000000} MSD (mm)} &
  {\color[HTML]{000000} 45.77±1.784*} &
  {\color[HTML]{000000} 0.925±0.401*} &
  {\color[HTML]{000000} 3.206±0.687*} &
  {\color[HTML]{000000} \textbf{0.356±0.198}} \\ \hline
{\color[HTML]{000000} } &
  {\color[HTML]{000000} DSC} &
  {\color[HTML]{000000} 00.00±00.00*} &
  {\color[HTML]{000000} 0.844±0.027*} &
  {\color[HTML]{000000} 00.00±00.00*} &
  {\color[HTML]{000000} \textbf{0.871±0.024}} \\
{\color[HTML]{000000} FC} &
  {\color[HTML]{000000} HD (mm)} &
  {\color[HTML]{000000} 160.8±5.453*} &
  {\color[HTML]{000000} 19.54±14.38*} &
  {\color[HTML]{000000} 160.8±5.453*} &
  {\color[HTML]{000000} \textbf{8.239±10.14}} \\
{\color[HTML]{000000} } &
  {\color[HTML]{000000} MSD (mm)} &
  {\color[HTML]{000000} 115.8±4.417*} &
  {\color[HTML]{000000} 0.41±0.104*} &
  {\color[HTML]{000000} 115.8±4.417*} &
  {\color[HTML]{000000} \textbf{0.321±0.074}} \\ \hline
{\color[HTML]{000000} } &
  {\color[HTML]{000000} DSC} &
  {\color[HTML]{000000} 0.957±0.014*} &
  {\color[HTML]{000000} 0.978±0.005*} &
  {\color[HTML]{000000} 0.767±0.046*} &
  {\color[HTML]{000000} \textbf{0.981±0.004}} \\
{\color[HTML]{000000} TB} &
  {\color[HTML]{000000} HD (mm)} &
  {\color[HTML]{000000} 69.56±15.75*} &
  {\color[HTML]{000000} 30.44±25.27*} &
  {\color[HTML]{000000} 82.32±10.55*} &
  {\color[HTML]{000000} \textbf{12.67±15.11}} \\
{\color[HTML]{000000} } &
  {\color[HTML]{000000} MSD (mm)} &
  {\color[HTML]{000000} 1.915±1.499*} &
  {\color[HTML]{000000} 0.403±0.277*} &
  {\color[HTML]{000000} 15.19±2.632*} &
  {\color[HTML]{000000} \textbf{0.332±0.092}} \\ \hline
{\color[HTML]{000000} } &
  {\color[HTML]{000000} DSC} &
  {\color[HTML]{000000} 0.251±0.015*} &
  {\color[HTML]{000000} 0.834±0.042*} &
  {\color[HTML]{000000} 0.306±0.075*} &
  {\color[HTML]{000000} \textbf{0.847±0.040}} \\
{\color[HTML]{000000} TC} &
  {\color[HTML]{000000} HD (mm)} &
  {\color[HTML]{000000} 111.1±8.961*} &
  {\color[HTML]{000000} 19.61±17.55*} &
  {\color[HTML]{000000} 86.39±6.187*} &
  {\color[HTML]{000000} \textbf{5.466±4.406}} \\
{\color[HTML]{000000} } &
  {\color[HTML]{000000} MSD (mm)} &
  {\color[HTML]{000000} 6.488±4.093*} &
  {\color[HTML]{000000} 0.322±0.111*} &
  {\color[HTML]{000000} 39.15±3.628*} &
  {\color[HTML]{000000} \textbf{0.287±0.103}} \\ \hline
\multicolumn{2}{c}{{\color[HTML]{000000} Parameter Number (million)}} &
  {\color[HTML]{000000} 0.076} &
  {\color[HTML]{000000} 0.037} &
  {\color[HTML]{000000} 121.5} &
  {\color[HTML]{000000} \textbf{0.167}} \\ \hline
\multicolumn{2}{c}{Training Speed (s/step)} &
  2.50 &
  1.74 &
  \textbf{1.67} &
  {\color[HTML]{000000} 1.73} \\ \hline
\multicolumn{2}{c}{Inference Speed GPU (s/step)} &
  2.77 &
  2.10 &
  \textbf{1.87} &
  2.02 \\ \hline
\multicolumn{2}{c}{Inference Speed CPU (s/step)} &
  8.3 &
  8.7 &
  12.2 &
  \textbf{5.5} \\ \hline
\end{tabular}}
\caption{OAI-ZIB knee dataset performance of models with their original losses under same training time and 12G VRAM graphic card. Eight hours of training is chosen as it is the time when CAN3D reaches peak performance. Parameter numbers and averaged computation speed of different network architectures are also listed. * represent metrics with p-value $<$0.05 (two-sided paired wilxocon signed-rank test) comparing to the CAN3D, Bold represent best score within each metric.}
\label{Table:zib_result}
\end{table*}

\subsection{Experiment III: Pelvis Dataset}

\begin{figure*}[!tp]
    \centering
    \includegraphics[width=0.9\textwidth]{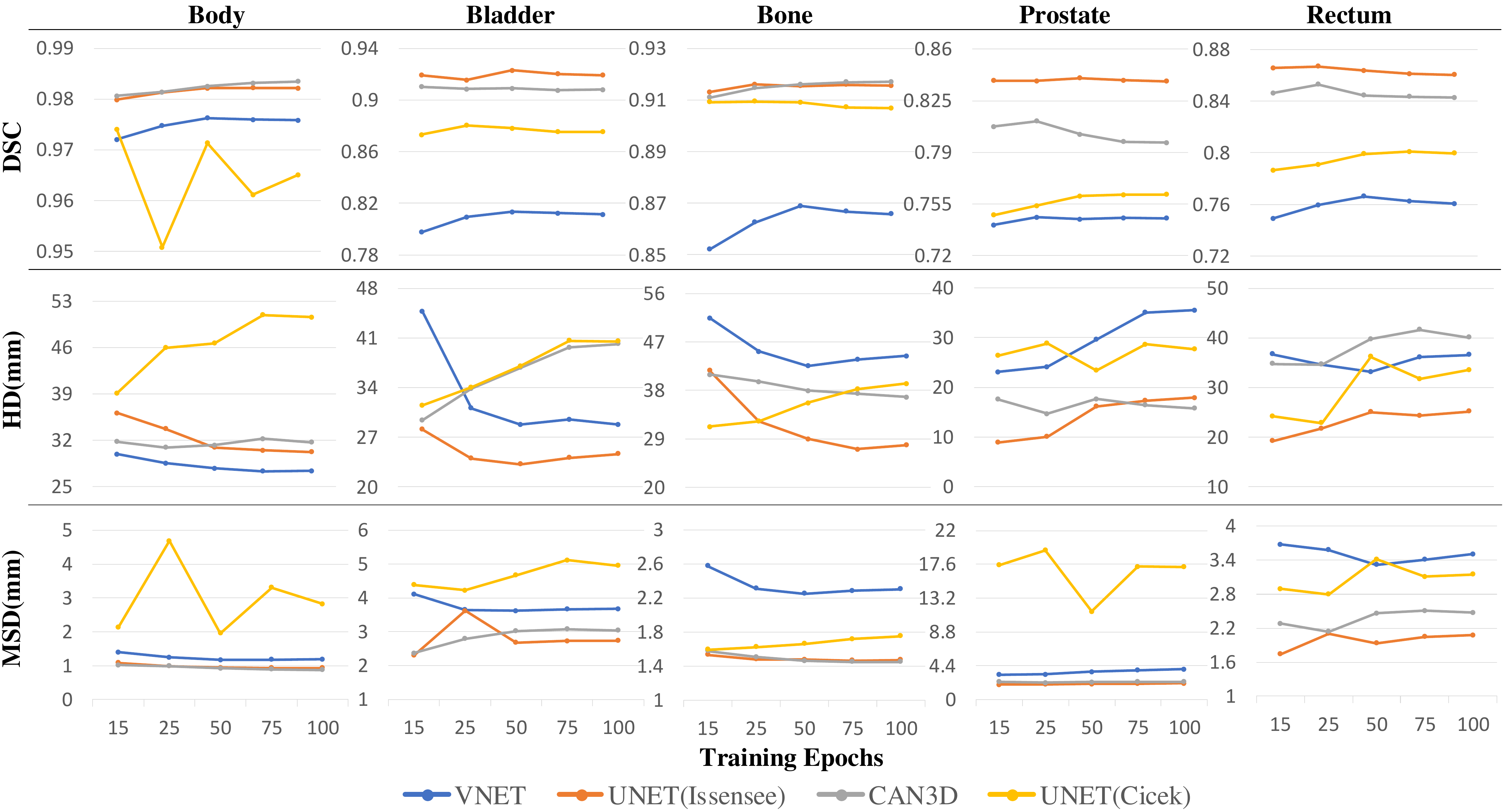}
    \caption{Pelvis dataset performance of VNET~\citep{milletari2016v}, UNET3D~\citep{cieck2016}, improved UNET3D~\citep{isensee2017brain} and CAN3D trained under 15, 25, 50, 75 and 100 epochs. Metrics such as Dice similarity coefficient (DSC), Hausdorff Distance(HD) and mean surface distance (MSD) for each class are plotted separately with single data point representing the averaged three-fold cross-validation results.}
    \label{fig:hip_plot}
\end{figure*}

\begin{table*}[!tp]
\centering
\scalebox{0.9}{
\begin{tabular}{cccccc}
\hline
{\color[HTML]{000000} Region} &
  {\color[HTML]{000000} Mean Metric} &
  {\color[HTML]{000000} UNET3D (Cicek)} &
  {\color[HTML]{000000} UNET3D (Issensee)} &
  {\color[HTML]{000000} VNET} &
  {\color[HTML]{000000} CAN3D} \\ \hline
{\color[HTML]{000000} } &
  {\color[HTML]{000000} DSC} &
  {\color[HTML]{000000} 0.948±0.044} &
  {\color[HTML]{000000} 0.980±0.011*} &
  {\color[HTML]{000000} 0.972±0.009*} &
  {\color[HTML]{000000} \textbf{0.981±0.009}} \\
{\color[HTML]{000000} Body} &
  {\color[HTML]{000000} HD (mm)} &
  {\color[HTML]{000000} 46.78±25.98} &
  {\color[HTML]{000000} 36.01±12.12*} &
  {\color[HTML]{000000} \textbf{28.99±9.070*}} &
  {\color[HTML]{000000} 30.89±11.21} \\
{\color[HTML]{000000} } &
  {\color[HTML]{000000} MSD (mm)} &
  {\color[HTML]{000000} 4.438±4.267*} &
  {\color[HTML]{000000} 1.079±0.590*} &
  {\color[HTML]{000000} 1.372±0.468*} &
  {\color[HTML]{000000} \textbf{0.990±0.486}} \\ \hline
{\color[HTML]{000000} } &
  {\color[HTML]{000000} DSC} &
  {\color[HTML]{000000} 0.855±0.173*} &
  {\color[HTML]{000000} \textbf{0.919±0.089*}} &
  {\color[HTML]{000000} 0.799±0.155*} &
  {\color[HTML]{000000} 0.909±0.108} \\
{\color[HTML]{000000} Bladder} &
  {\color[HTML]{000000} HD (mm)} &
  {\color[HTML]{000000} 32.18±39.26} &
  {\color[HTML]{000000} \textbf{28.08±36.07*}} &
  {\color[HTML]{000000} 40.29±30.98} &
  {\color[HTML]{000000} 33.88±38.09} \\
{\color[HTML]{000000} } &
  {\color[HTML]{000000} MSD (mm)} &
  {\color[HTML]{000000} 5.276±9.552*} &
  {\color[HTML]{000000} \textbf{2.301±3.540*}} &
  {\color[HTML]{000000} 3.997±3.007*} &
  {\color[HTML]{000000} 2.789±4.381} \\ \hline
{\color[HTML]{000000} } &
  {\color[HTML]{000000} DSC} &
  {\color[HTML]{000000} 0.907±0.021*} &
  {\color[HTML]{000000} 0.913±0.020*} &
  {\color[HTML]{000000} 0.849±0.033*} &
  {\color[HTML]{000000} \textbf{0.915±0.019}} \\
{\color[HTML]{000000} Bone} &
  {\color[HTML]{000000} HD (mm)} &
  {\color[HTML]{000000} \textbf{36.23±22.81*}} &
  {\color[HTML]{000000} 41.73±28.95} &
  {\color[HTML]{000000} 52.22±16.87} &
  {\color[HTML]{000000} 39.66±24.71} \\
{\color[HTML]{000000} } &
  {\color[HTML]{000000} MSD (mm)} &
  {\color[HTML]{000000} 1.653±0.427*} &
  {\color[HTML]{000000} 1.533±0.391*} &
  {\color[HTML]{000000} 2.632±0.910*} &
  {\color[HTML]{000000} \textbf{1.505±0.383}} \\ \hline
{\color[HTML]{000000} } &
  {\color[HTML]{000000} DSC} &
  {\color[HTML]{000000} 0.779±0.101*} &
  {\color[HTML]{000000} \textbf{0.866±0.050}} &
  {\color[HTML]{000000} 0.756±0.092*} &
  {\color[HTML]{000000} 0.853±0.053} \\
{\color[HTML]{000000} Rectum} &
  {\color[HTML]{000000} HD (mm)} &
  {\color[HTML]{000000} 27.12±26.61*} &
  {\color[HTML]{000000} \textbf{19.18±25.00}} &
  {\color[HTML]{000000} 42.48±35.33} &
  {\color[HTML]{000000} 34.65±36.54} \\
{\color[HTML]{000000} } &
  {\color[HTML]{000000} MSD (mm)} &
  {\color[HTML]{000000} 3.221±2.837*} &
  {\color[HTML]{000000} \textbf{1.739±1.052*}} &
  {\color[HTML]{000000} 3.921±4.382*} &
  {\color[HTML]{000000} 2.138±1.369} \\ \hline
\multicolumn{1}{l}{} &
  DSC &
  0.735±0.158* &
  {\color[HTML]{000000} \textbf{0.839±0.069*}} &
  0.735±0.127* &
  0.811±0.090 \\
\multicolumn{1}{l}{Prostate} &
  HD (mm) &
  31.99±54.62 &
  {\color[HTML]{000000} \textbf{8.890±8.867*}} &
  28.53±22.21* &
  14.63±16.30 \\
\multicolumn{1}{l}{} &
  MSD (mm) &
  19.76±41.76* &
  {\color[HTML]{000000} \textbf{1.920±0.771*}} &
  3.679±2.704* &
  2.187±0.986 \\ \hline
\multicolumn{2}{c}{{\color[HTML]{000000} Parameter Number (million)}} &
  {\color[HTML]{000000} 1.196} &
  {\color[HTML]{000000} 2.369} &
  {\color[HTML]{000000} 264.9} &
  {\color[HTML]{000000} \textbf{0.171}} \\ \hline
\multicolumn{2}{c}{Training Speed (s/step)} &
  1.57 &
  1.15 &
  1.28 &
  {\color[HTML]{000000} \textbf{0.8}} \\ \hline
\multicolumn{2}{c}{Inference Speed GPU (s/step)} &
  1.13 &
  0.92 &
  0.76 &
  \textbf{0.76} \\ \hline
\multicolumn{2}{c}{Inference Speed CPU (s/step)} &
  5.4 &
  5.0 &
  6.2 &
  \textbf{3.2} \\ \hline
\end{tabular}}
\caption{Pelvis dataset performance of models with their original losses under same training time and 12G VRAM graphic card. Three hours of training is chosen as it is the time when CAN3D reaches peak performance. Parameter numbers and averaged computation speed of different network architectures are also listed. * represent metrics with p-value $<$0.05 (two-sided paired Wilxocon signed-rank test) comparing to the CAN3D, Bold represent best score within each metric.}
\label{Table:hip_result}
\end{table*}

\begin{table*}[!tp]
    \centering
    \scalebox{0.9}{
    \begin{tabular}{cccccccccc}
    \hline
    Method &
    \begin{tabular}[c]{@{}c@{}}Body \\ median \\ DSC\end{tabular} &
    \begin{tabular}[c]{@{}c@{}}Bone\\ median\\  DSC\end{tabular} &
    \begin{tabular}[c]{@{}c@{}}Bladder \\ median \\ DSC\end{tabular} &
    \begin{tabular}[c]{@{}c@{}}Rectum \\ median \\ DSC\end{tabular} &
    \begin{tabular}[c]{@{}c@{}}Prostate \\ median \\ DSC\end{tabular} &
    \begin{tabular}[c]{@{}c@{}}Prostate \\ median MSD\\  (mm)\end{tabular} &
    \begin{tabular}[c]{@{}c@{}}Prostate \\ median HD \\ (mm)\end{tabular} &
    \begin{tabular}[c]{@{}c@{}}Prostate \\ mean \\ DSC\end{tabular} &
    \begin{tabular}[c]{@{}c@{}}Inference \\ time (s)\end{tabular} \\ \hline
    Dowling et al &  \textcolor{red}{\textbf{1}} & 0.92 & 0.86          & 0.85          & 0.82          & 2.04          & 13.3          & 0.80          & 7200           \\
    Chandra et al & 0.94       & 0.81          & 0.87          & 0.79          & 0.81          & 2.08          &  \textcolor{red}{\textbf{9.60}} & 0.79          & 720            \\
    CAN3D         & 0.99       &   \textcolor{red}{\textbf{0.92}}          &  \textcolor{red}{\textbf{0.96}} &  \textcolor{red}{\textbf{0.86}} &  \textcolor{red}{\textbf{0.84}} &  \textcolor{red}{\textbf{1.92}} & 10.61          &  \textcolor{red}{\textbf{0.81}} &  \textcolor{red}{\textbf{3.2}} \\ \hline
    \end{tabular}
    }
    \caption{Performance of the proposed \ac{CAN3D}, a multi-object weighted deformable model proposed by \citet{Chandra2016a}. and a Multi-atlas approach proposed by \citet{Dowling2015}. }
    \label{Table:ML}
\end{table*}

Figure~\ref{fig:hip_plot} compares the performance of all models by depicting different class accuracy under various epochs. For peak performance, both CAN3D and improved U-Net3D attain comparable \ac{DSC} scores for all classes (except prostate), which are much higher than V-Net and U-Net3D's. As for \ac{HD}, the CAN3D only provides similar performance to improved U-Net3D for body and prostate, and worse results for the rest of the classes than other models. Despite having undesirable performance in terms of \ac{HD}, CAN3D again achieves indistinguishable \ac{MSD} performance to the improved U-Net3D for all classes, which are much better than other models.

The quantitative comparison of models trained under the same three hours session is recorded in Table~\ref{Table:hip_result}. The table indicates that CAN3D and improved U-Net3D achieve similar results with a small disparity in DSC, surpassing U-Net3D and V-Net by a considerable margin. While achieving better surface distance than improved U-Net3D for rigid components such as body and bone, the CAN3D fails to compete with improved U-Net3D on non-rigid components such as bladder, rectum and prostate. Hence, CAN3D generally performs better than the U-Net3D and V-Net, however, its performance does not match the improved U-Net3D in terms of segmentation accuracy. Table~\ref{Table:hip_result} also tabulates the parameter numbers and computational speed for the chosen models. The \ac{CAN3D} only uses 0.17M parameters which are about 14\% of the parameters used by U-Net~\citep{cieck2016}, 7\% used by improved U-Net~\citep{isensee2017brain} and 0.07\% used by V-Net~\citep{milletari2016v}. Moreover, \ac{CAN3D}'s training time is about 1.5 to 2 times faster for each step than other models. While the advantage of the proposed model in terms of inference time is not significant when testing using GPU, the superiority of CAN3D's compact structure becomes prominent when swapping to CPU as it only uses 3.2 s/step during inference which is about 40\% to 50\% less than other models.

The supplementary results for traditional automatic methods are also presented in Table~\ref{Table:ML}. Our proposed method reaches score of (0.96, 0.86, 0.84) and (0.81) in terms of median \ac{DSC} for (bladder, rectum, prostate) and mean \ac{DSC} for (prostate). This is a notable improvement compared with previous work, especially for the bladder, whose accuracy increased by around 10\%. However, no distinct advantages are found for the proposed method in terms of the rigid structures (body and bone) comparing to the multi-atlas approach~\citep{Dowling2015}. As for surface accuracy, a tie is observed between our model and the multi-object deformable model~\citep{Chandra2016a}, where the proposed model has a clear advantage for \ac{MSD} while the deformable model has a better score for \ac{HD}. A breakthrough is also observed in computation times for our proposed model where its inference time on CPU is several orders of magnitude faster than other automatic methods.

\begin{figure*}[!t]
    \centering
    \includegraphics[width=0.8\textwidth]{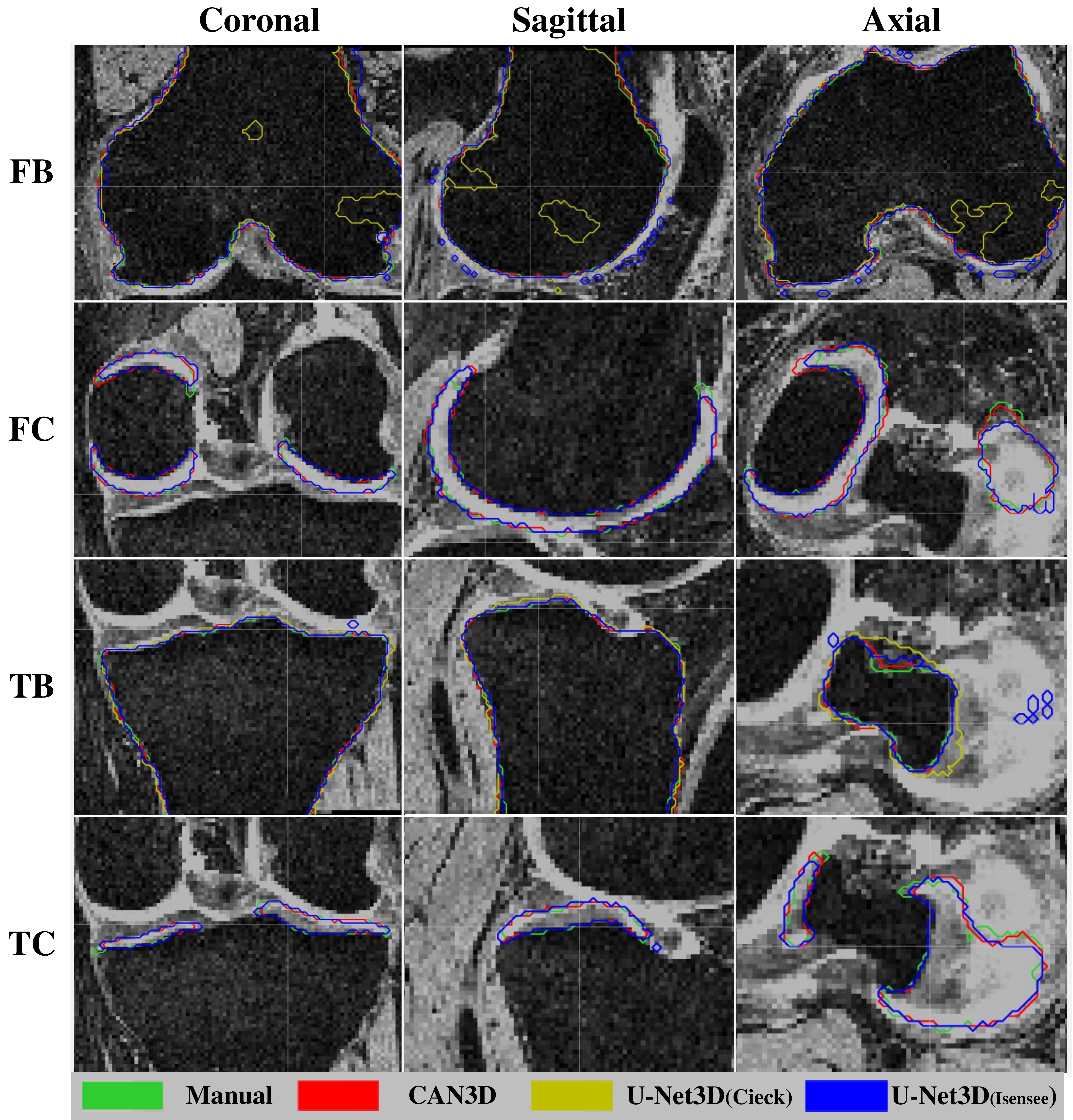}
    \caption{Extracted slices of segmented results for OAI-ZIB dataset. Each row shows a patient's data from three different views (coronal, sagittal and axial). All classes(Femoral Bone (FB),Tibial Bone (TB), Femoral Cartilage (FC) and Tibial Cartilage (TC)) and all models except V-net is shown as V-Net completely failed for this experiment.}
    \label{fig:zib_visual}
\end{figure*}

\begin{figure*}[!t]
    \centering
    \includegraphics[width=0.8\textwidth]{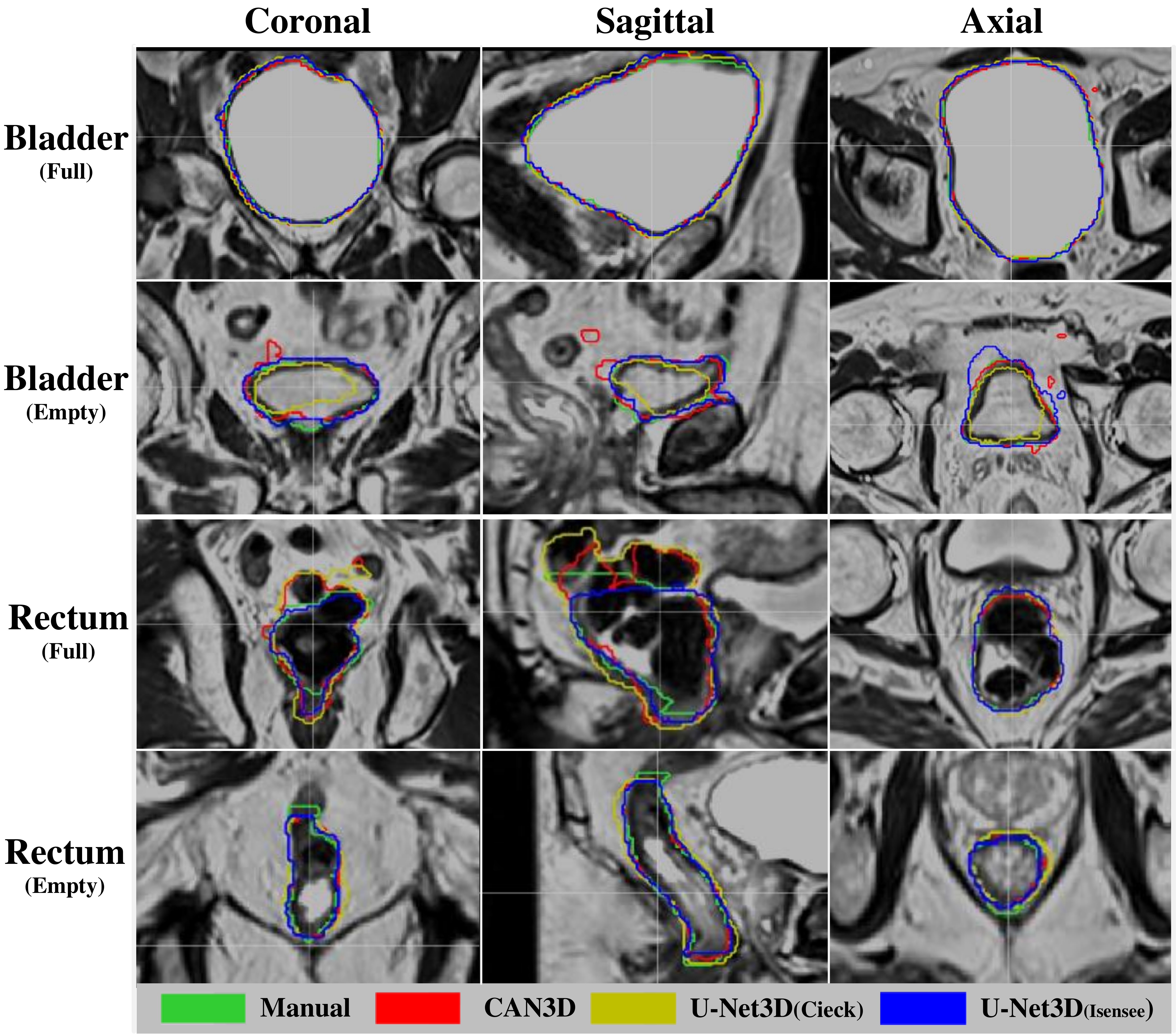}
    \caption{Extracted slices of segmented results for pelvis dataset. Each row shows the image data of a patient from three different views (coronal, sagittal and axial). Bladder and rectum classes are chosen for display as they have the worst surface distance for CAN3D. (first, third) and (second, fourth) rows show the patient with the full and empty bladder and rectum respectively. All models except V-net is shown as V-Net failed for this experiment.}
    \label{fig:hip_visual}
\end{figure*}

\section{Discussion}
In this study, we proposed a CNN model with an extremely compact structure to semantically segment large \ac{3D} MRI volumes efficiently under limited resources. The network is built based on a compact context aggregation module (\ac{CAM}) to minimize the number of parameters and down-sampling operations required while sustaining the same receptive field size. A novel \ac{DSF} loss function was also presented to optimize the proposed model by restraining both volumetric and surface accuracy during the training. We evaluated our proposed model and loss function using 3-fold cross-validation on both private pelvis dataset and open OAI-ZIB knee dataset.

\subsection{Dice Squared Focal Loss}
Table~\ref{Table:loss} highlights our proposed \ac{DSL}'s superiority using a compact network structure under a fast training process. It achieves superior results for both volume-based and surface-based metrics than traditional losses such as \ac{WCE} and \ac{DSC}. Compared to the original \ac{DSC} loss, the \ac{DSL}, which is extended from it, successfully improves the final segmentation accuracy with even faster computation speed. When we compare \ac{DSF} to \ac{DSL}, although no significant improvement is found for metrics such as \ac{DSC} and \ac{MSD}, the extra focal loss component within the \ac{DSF} significantly improves the \ac{HD} with minimal burden to computation speed. This shows that the extra Focal Loss component successfully preserves more boundary details, thus improving the surface precision of final segmentation. Overall, the proposed \ac{DSF} improves the volume-based accuracy compared to \ac{DSC}, and generates better surface-based results than the \ac{CE}.

Table~\ref{Table:loss_other_model} further demonstrates the advantage of \ac{DSF} as it out performs weighted \ac{CE} in almost all measurements under U-Net3D. However, the change brought by \ac{DSF} to Improved U-Net3D is not as significant that its performance is even worse than the weighted \ac{DSC} for extremely imbalanced classes such as rectum and prostate. Since the Improved U-Net3D's structure is more complicated than the original U-Net3D, this result suggests the proposed \ac{DSF} performs the best for compact model with fewer parameters. This assumption is also supported by the results from Table~\ref{Table:loss}, as \ac{DSF} achieves much better performance than weighted \ac{DSC} for our compact CAN3D.

\subsection{CAN3D}

\begin{figure}[!t]
    \centering
    \includegraphics[width=0.4\textwidth]{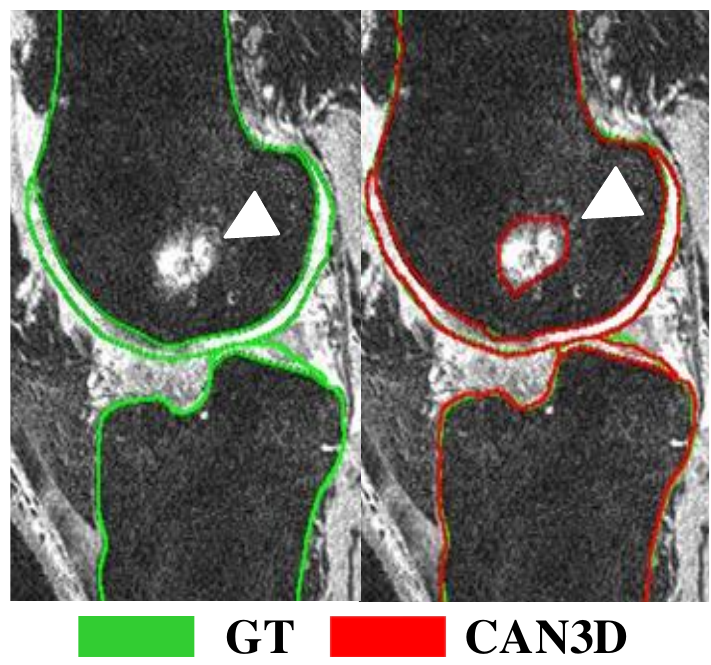}
    \caption{CAN3D (red) is shown to detect bone marrow lesions, which are not outlined in ground truth (green) }
    \label{fig:lesion_zib}
\end{figure}

Based on the results shown for both knee and pelvis experiments, traditional networks such as U-Net3D and V-Net do not compare favourably with newer architectures such as Improved U-Net3D and CAN3D. Both of them  fail completely for the knee dataset as their model parameters need to be truncated dramatically from their original design, so the larger input such as knee image (1.5 times larger than pelvis' in terms of resolution) can be propagated through the model during training. If we compare the parameter number of each model between pelvis and knee experiments respectively (shown in Table~\ref{Table:hip_result} and Table~\ref{Table:zib_result}), it is shown that the number of parameters plunges dramatically for all models except our CAN3D. While CAN3D holds approximately the same amount of parameters (98\%), this number is declined by 93\% for U-Net3D, 98\% for improved U-Net3D and 55\% for V-Net. As for Improved U-Net3D, even after losing a large portion of parameters, it could still provide accurate results with decent converging speed. However, instead of a close match with CAN3D during pelvis experiment, it is outperformed by CAN3D in every measurements during knee experiment, especially for \ac{HD}. Thus, it can be concluded that the performance of Improved U-Net3D is also capped severely after pruning its network structure for large input. On the other hand, the CAN3D could still preserve most of the parameters and run at full capacity regardless the increasing size of the input, which shows the advantage of CAN3D's compact structure that it only consumes a small portion of resources during computation and could efficiently utilise a limited amount of parameters for the state-of-art performance.

The CAN3D also shows its efficiency in promising computation speeds for both experiments. For the pelvis experiment, CAN3D's GPU training and CPU inference speed are about (30\% to 50\%) and (40\% to 50\%) faster than other models respectively, thanks to its extreme compact structure with only (0.07\% to 14\%) parameters used by others. For the knee experiment, because of the trade-off between models' complexity and their performance, CAN3D ends up having relatively more parameters than other U-Nets. However, it still out performs the CPU inference speed which is about (35\% to 55\%) faster despite having more parameters. This is because of the fewer convolution and sub-sampling operations involved through model for CAN3D, thus, faster inference could be achieved with fewer computations. Since GPU has ability to run multiple computations in parallel, the inference speed advantage is emphasised more when running using CPU. Moreover, the reason for V-Net having faster GPU computation speed with more parameters is because most of the parameters exist at the deep sub-sampled layers, thus, each computation can be computed quickly but has less effect to the final accuracy. This is further supported by its much longer inference speed under CPU when all computations are done sequentially.

Figure~\ref{fig:zib_visual} visualises the typical extracted slices of segmented results for the OAI-ZIB knee dataset. The V-Net's results are excluded from the figure as it could not produce acceptable segmentations because it had to be truncated to fit into memory. As shown in the figure, U-Net3D (Yellow) fails to generate accurate segmentation for both cartilages and a large portion of \ac{FB}, which explains its poor performance when convergence is insufficient. The improved U-Net3D, while producing some reasonable segmentations, is also shown to generate false-positive prediction for pixels that are distant from the target, which explains its close \ac{DSC} to CAN3D, but worse results for \ac{HD}. Surprisingly,  CAN3D is also shown to detect bone marrow lesions within \ac{FB} in Figure~\ref{fig:lesion_zib}, which is not outlined in the ground truth. This suggests that CAN3D has promise of detecting abnormalities within the healthy tissues thanks to its large receptive fields and minimal resolution losses during the training. A potential future work could be to generate a pipeline that can detect abnormalities and produce more robust segmentation to handle pathologies.

Figure~\ref{fig:hip_visual} shows the visualization of pelvic results of all models under different scenarios. For bladders, all models achieve good performance when the bladders are full with large shape. However, the models' accuracy decreases dramatically as bladders become empty with more irregular shapes. For instance, the U-Net3D fails to segment partial bladder from all viewpoints, and both improved U-Net3D and CAN3D generate false-positive errors around the target. Compared to the improved U-Net3D, CAN3D is more likely to generate outliers far away from the ground truth. As for the empty rectums, all models again generate similar results with high accuracy. However, when rectums are full with their features such as contrast or density closely resemble the surrounding tissues, the CAN3D begins to incorrectly segment the targets' surrounding tissues that are not part of the ground truth label. Both scenarios for either bladder or rectum explain the poor \ac{HD} but much better \ac{MSD} results for CAN3D in the pelvis experiment and reveal the possible limitation of CAN3D's compact structure when segmenting imbalanced objects under a more complex environment. Simple solutions such as the post-processing techniques mentioned regarding the knee experiment could be easily implemented to overcome the distant outlier problem, and the outcomes have been proven very positive in Table~\ref{Table:knee_32g}.

Overall, CAN3D's efficiency is much higher than other models as it accomplishes the state-of-art performance with much fewer parameters and faster computational time. The small memory footprint of its model ensure that its performance is barely affected by the size of the input under limited resources. Even though CAN3D lacks performance in constraining the maximum surface error on the contour of imbalanced classes, it can still produce the best overall surface distance in an averaging term. It can also  predict a single image with half of the CPU time used by other CNN models and couple orders of magnitude faster than complex pipeline such as \citet{ambellan2019automated}.  A clear advantage can also be observed for CAN3D in Figure~\ref{Table:ML} when compared to other traditional automatic methods, which emphasize the advantage of the compact structure of our proposed model for clinical deployment where performance is critical.

\subsection{Future work}
Although experiment results have demonstrated that the proposed model could achieve accurate and fast semantic segmentation for large input images under restricted resources, there are still some limitations where segmented results might not be satisfactory in a clinical situation. While our model could provide high volumetric accuracy with precise surface distance in general, its constraints for extreme errors deviated from target surface still need to be improved. Fortunately, since the generated outliers are relatively small and disconnected from ground truth segmentation, post-processing techniques can easily be implemented to remove them and improve \ac{HD} performance. The model also needs to improve its ability to distinguish the target from its surrounding tissues with similar features. Moreover, thanks to its compact structure, various possible solutions can be easily implemented as there will be no concerns of hardware limitation. In future work, we aim to take the advantage of this compact structure and implement it with more advanced deep learning approach such as generative adversarial network, to further enhance the surface accuracy and make it possible to accelerate high-quality segmentation for the clinical setting.

\section{Conclusion}
We proposed a compact deep neural network, called \ac{CAN3D}, for efficient semantic segmentation in \ac{3D} MRI. The designed model utilized context aggregation to preserve sizeable receptive fields while minimizing the required number of down-sampling operations. It was shown to effectively improve final segmentation accuracy by reducing the loss of resolution and fully exploiting the 3D spatial information during training. It achieved state-of-art performance with orders of magnitude fewer parameters and multiple times faster computation time than other 3D networks, particularly when segmenting data with extremely large shapes under limited hardware constraints. The proposed network also outperformed other traditional automatic methods and more complex CNN-based pipelines with a dramatic advance in inference time. Moreover, a loss function called \ac{DSF} was also designed to optimize the training process by restraining both volumetric and surface difference with no extra cost. Our experiments demonstrated the promise of this novel loss in improving the segmentation performance and accelerating the convergence during training.
\acrodef{3D}{\emph{three dimensional}}
\acrodef{CNN}{\emph{Convolutional Neural Network}}
\acrodef{MR}{\emph{Magnetic Resonance}}
\acrodef{DSC}{\emph{Dice Similarity Coefficient}}
\acrodef{MSD}{\emph{Mean Surface Distance}}
\acrodef{HD}{\emph{Hausdorff Distance}}
\acrodef{SIMPLE}{\emph{Selective and Iterative Method for Performance Level Estimation}}
\acrodef{CT}{\emph{Computed Tomography}}
\acrodef{DNN}{\emph{Deep neural networks}}
\acrodef{2D}{\emph{two dimensional}}
\acrodef{WCE}{\emph{Weighted Cross-Entropy}}
\acrodef{SSM}{\emph{Statistical Shape Models}}
\acrodef{CNN}{\emph{Convolutional Neural Network}}
\acrodef{GPUs}{\emph{Graphics Processing Units}}
\acrodef{HPC}{\emph{high-performance computing}}
\acrodef{CPUs}{\emph{Central Processing Units}}
\acrodef{CAN3D}{\emph{Context Aggregation Network 3D}}
\acrodef{DSF}{\emph{Dice Squared Focal Loss}}
\acrodef{DSL}{\emph{Dice Squared Loss}}
\acrodef{RF}{\emph{receptive field}}
\acrodef{CAN}{\emph{Context Aggregation Network}}
\acrodef{LReLU}{\emph{Leaky ReLU}}
\acrodef{ABN}{\emph{Adaptive Batch Normalization}}
\acrodef{BN}{\emph{Batch Normalization}}
\acrodef{FL}{\emph{Focal Loss}}
\acrodef{MSE}{\emph{Mean Square Error}}
\acrodef{CE}{\emph{Cross-entropy}}
\acrodef{SPACE}{\emph{Sampling Perfection with Application optimized Contrasts using different flip angle Evolution}}
\acrodef{COCAN3D}{\emph{compact context aggregation network 3D}}
\acrodef{OAI}{\emph{Osteoarthritis Initiative}}
\acrodef{ZIB}{\emph{Zuse Institute Berlin}}
\acrodef{FB}{\emph{Femoral Bone}}
\acrodef{FC}{\emph{Femoral Cartilage}}
\acrodef{TB}{\emph{Tibial Bone}}
\acrodef{TC}{\emph{Tibial Cartilage}}
\acrodef{VRAM}{\emph{Video Memory}}
\acrodef{CAM}{\emph{Context Aggregation Module}}
\acrodef{AdaIN}{\emph{adaptive instance normalization}}
\acrodef{IN}{\emph{instance normalization}}
\acrodef{OAI}{\emph{Osteoarthritis Initiative}}
\acrodef{FCN}{\emph{Fully Convolutional Network}}
\acrodef{ASPP}{\emph{Atrous Spatial Pyramid Pooling}}

\bibliographystyle{unsrtnat}
\bibliography{references}  

\begin{thebibliography}{68}
\providecommand{\natexlab}[1]{#1}
\providecommand{\url}[1]{\texttt{#1}}
\expandafter\ifx\csname urlstyle\endcsname\relax
  \providecommand{\doi}[1]{doi: #1}\else
  \providecommand{\doi}{doi: \begingroup \urlstyle{rm}\Url}\fi

\bibitem[Nyholm and Jonsson(2014)]{Nyholm2014}
T.~Nyholm and J.~Jonsson.
\newblock Counterpoint: {opportunities} and {challenges} of a {magnetic}
  {resonance} {imaging}-{only} {radiotherapy} {work} {flow}.
\newblock \emph{Seminars in Radiation Oncology}, 24\penalty0 (3):\penalty0
  175--180, July 2014.
\newblock ISSN 1053-4296.
\newblock \doi{10.1016/j.semradonc.2014.02.005}.

\bibitem[Pollard et~al.(2008)Pollard, Gwilym, and Carr]{pollard2008assessment}
T.~C.~B. Pollard, S.~E. Gwilym, and A.~J. Carr.
\newblock The assessment of early osteoarthritis.
\newblock \emph{The Journal of bone and joint surgery. British volume},
  90\penalty0 (4):\penalty0 411--421, 2008.

\bibitem[Pereira et~al.(2016)Pereira, Pinto, Alves, and Silva]{Pereira2016}
S.~Pereira, A.~Pinto, V.~Alves, and C.~A. Silva.
\newblock Brain tumor segmentation using convolutional neural networks in {MRI}
  images.
\newblock \emph{IEEE Transactions on Medical Imaging}, 35\penalty0
  (5):\penalty0 1240--1251, 2016.
\newblock ISSN 0278-0062.
\newblock \doi{10.1109/Tmi.2016.2538465}.

\bibitem[Yu and Koltun(2015)]{yu2015CAN}
F.~Yu and V.~Koltun.
\newblock Multi-scale context aggregation by dilated convolutions.
\newblock \emph{arXiv preprint arXiv:1511.07122}, 2015.

\bibitem[Lin et~al.(2017)Lin, Goyal, Girshick, He, and
  Doll{\'a}r]{lin2017focal}
T.~Lin, P.~Goyal, R.~Girshick, K.~He, and P.~Doll{\'a}r.
\newblock Focal loss for dense object detection.
\newblock In \emph{Proceedings of the IEEE international conference on computer
  vision}, pages 2980--2988, 2017.

\bibitem[Shi et~al.(2018)Shi, Zeng, Zhang, Zhuang, Li, Yang, and
  Zheng]{shi2018bayesian}
Z.~Shi, G.~Zeng, L.~Zhang, X.~Zhuang, L.~Li, G.~Yang, and G.~Zheng.
\newblock Bayesian voxdrn: A probabilistic deep voxelwise dilated residual
  network for whole heart segmentation from {3D} {MR} images.
\newblock In \emph{International Conference on Medical Image Computing and
  Computer-Assisted Intervention}, pages 569--577. Springer, 2018.

\bibitem[Dowling and Greer(2021)]{pelvisdata2021}
J.~Dowling and P.~Greer.
\newblock Labelled weekly {MR} images of the male pelvis, 2021.
\newblock v1. CSIRO. Data Collection. \url{https://doi.org/10.25919/45t8-p065}.

\bibitem[Ambellan et~al.(2019)Ambellan, Tack, Ehlke, and
  Zachow]{ambellan2019automated}
F.~Ambellan, A.~Tack, M.~Ehlke, and S.~Zachow.
\newblock Automated segmentation of knee bone and cartilage combining
  statistical shape knowledge and convolutional neural networks: Data from the
  osteoarthritis initiative.
\newblock \emph{Medical Image Analysis}, 52:\penalty0 109--118, 2019.

\bibitem[Sharma and Ray(2006)]{sharma2006computer}
N.~Sharma and A.~K. Ray.
\newblock Computer aided segmentation of medical images based on hybridized
  approach of edge and region based techniques.
\newblock In \emph{Proc. Int. Conf. Math. Biol}, pages 150--155, 2006.

\bibitem[Ramesh et~al.(1995)Ramesh, Yoo, and Sethi]{ramesh1995thresholding}
N.~Ramesh, J.~Yoo, and I.~K. Sethi.
\newblock Thresholding based on histogram approximation.
\newblock \emph{IEEE Proceedings-Vision, Image and Signal Processing},
  142\penalty0 (5):\penalty0 271--279, 1995.

\bibitem[Wang et~al.(1996)Wang, Guerriero, and De~Sario]{wang1996comparison}
Z.~Wang, A.~Guerriero, and M.~De~Sario.
\newblock Comparison of several approaches for the segmentation of texture
  images.
\newblock \emph{Pattern Recognition Letters}, 17\penalty0 (5):\penalty0
  509--521, 1996.

\bibitem[Ghose et~al.(2012)Ghose, Oliver, Mart{\'\i}, Llad{\'o}, Vilanova,
  Freixenet, Mitra, Sidib{\'e}, and Meriaudeau]{ghose2012survey}
S.~Ghose, A.~Oliver, R.~Mart{\'\i}, X.~Llad{\'o}, J.~C. Vilanova, J.~Freixenet,
  J.~Mitra, D.~Sidib{\'e}, and F.~Meriaudeau.
\newblock A survey of prostate segmentation methodologies in ultrasound,
  magnetic resonance and computed tomography images.
\newblock \emph{Computer methods and programs in biomedicine}, 108\penalty0
  (1):\penalty0 262--287, 2012.

\bibitem[Ebrahimkhani et~al.(2020)Ebrahimkhani, Jaward, Cicuttini, Dharmaratne,
  Wang, and de~Herrera]{ebrahimkhani2020review}
S.~Ebrahimkhani, M.~H. Jaward, F.~M. Cicuttini, A.~Dharmaratne, Y.~Wang, and
  A.~G.~S. de~Herrera.
\newblock A review on segmentation of knee articular cartilage: from
  conventional methods towards deep learning.
\newblock \emph{Artificial Intelligence in Medicine}, page 101851, 2020.

\bibitem[Withey and Koles(2007)]{withey2007three}
D.~J. Withey and Z.~J. Koles.
\newblock Three generations of medical image segmentation: Methods and
  available software.
\newblock \emph{International Journal of Bioelectromagnetism}, 9\penalty0
  (2):\penalty0 67--68, 2007.

\bibitem[Klein et~al.(2008)Klein, Heide, Lips, Vulpen, Staring, and
  Pluim]{Klein2008}
S.~Klein, U.~A. Heide, I.~M. Lips, M.~Vulpen, M.~Staring, and J.~P.~W. Pluim.
\newblock Automatic segmentation of the prostate in {3D} {MR} images by atlas
  matching using localized mutual information.
\newblock \emph{Medical Physics}, 35\penalty0 (4):\penalty0 1407--1417, 2008.
\newblock \doi{10.1118/1.2842076}.

\bibitem[Shiee et~al.(2010)Shiee, Bazin, Ozturk, Reich, Calabresi, and
  Pham]{shiee2010topology}
N.~Shiee, P.~Bazin, A.~Ozturk, D.~S. Reich, P.~A. Calabresi, and D.~L. Pham.
\newblock A topology-preserving approach to the segmentation of brain images
  with multiple sclerosis lesions.
\newblock \emph{NeuroImage}, 49\penalty0 (2):\penalty0 1524--1535, 2010.

\bibitem[Tamez-Pe{\~n}a et~al.(2011)Tamez-Pe{\~n}a, Gonz{\'a}lez, Farber, Baum,
  Schreyer, and Totterman]{tamez2011atlas}
J.~Tamez-Pe{\~n}a, P.~Gonz{\'a}lez, J.~Farber, K.~Baum, E.~Schreyer, and
  S.~Totterman.
\newblock Atlas based method for the automated segmentation and quantification
  of knee features: Data from the osteoarthritis initiative.
\newblock In \emph{2011 IEEE International Symposium on Biomedical Imaging:
  From Nano to Macro}, pages 1484--1487. IEEE, 2011.

\bibitem[Dowling et~al.(2015{\natexlab{a}})Dowling, Sun, Pichler,
  Rivest-H\'{e}nault, Ghose, Richardson, Wratten, Martin, Arm, Best, Chandra,
  Fripp, Menk, and Greer]{Dowling2015}
J.~A. Dowling, J.~Sun, P.~Pichler, D.~Rivest-H\'{e}nault, S.~Ghose,
  H.~Richardson, C.~Wratten, J.~Martin, J.~Arm, L.~Best, S.~S. Chandra,
  J.~Fripp, F.~W. Menk, and P.~B. Greer.
\newblock Automatic substitute {CT} generation and contouring for {MRI}-alone
  external beam radiation therapy from standard {MRI} sequences.
\newblock \emph{International Journal of Radiation Oncology * Biology *
  Physics}, 93\penalty0 (5):\penalty0 1144--1153, 2015{\natexlab{a}}.
\newblock ISSN 0360-3016.
\newblock \doi{10.1016/j.ijrobp.2015.08.045}.

\bibitem[Schmid et~al.(2011)Schmid, Kim, and {Magnenat-Thalmann}]{Schmid2011}
J.~Schmid, J.~Kim, and N.~{Magnenat-Thalmann}.
\newblock Robust statistical shape models for {MRI} bone segmentation in
  presence of small field of view.
\newblock \emph{Medical Image Analysis}, 15\penalty0 (1):\penalty0 155-- 168,
  2011.
\newblock ISSN 1361-8415.
\newblock \doi{DOI: 10.1016/j.media.2010.09.001}.

\bibitem[Chandra et~al.(2014)Chandra, Xia, Engstrom, Crozier, Schwarz, and
  Fripp]{Chandra2014a}
S.~S. Chandra, Y.~Xia, C.~Engstrom, S.~Crozier, R.~Schwarz, and J.~Fripp.
\newblock Focused shape models for hip joint segmentation in {3D} magnetic
  resonance images.
\newblock \emph{Medical Image Analysis}, 18\penalty0 (3):\penalty0 567 -- 578,
  2014.
\newblock ISSN 1361-8415.
\newblock \doi{10.1016/j.media.2014.02.002}.

\bibitem[Engstrom et~al.(2011)Engstrom, Fripp, Jurcak, Walker, Salvado, and
  Crozier]{Engstrom2011}
C.~M. Engstrom, J.~Fripp, V.~Jurcak, D.~G. Walker, O.~Salvado, and S.~Crozier.
\newblock Segmentation of the quadratus lumborum muscle using statistical shape
  modeling.
\newblock \emph{Journal of Magnetic Resonance Imaging}, 33\penalty0
  (6):\penalty0 1422--1429, June 2011.
\newblock ISSN 1522-2586.
\newblock \doi{10.1002/jmri.22188}.

\bibitem[Martin et~al.(2010)Martin, Troccaz, and Daanen]{Martin2010}
S.~Martin, J.~Troccaz, and V.~Daanen.
\newblock Automated segmentation of the prostate in {3D} {MR} images using a
  probabilistic atlas and a spatially constrained deformable model.
\newblock \emph{Medical Physics}, 37\penalty0 (4):\penalty0 1579--1590, 2010.
\newblock \doi{10.1118/1.3315367}.

\bibitem[Ecabert et~al.(2008)Ecabert, Peters, Schramm, Lorenz, Berg, Walker,
  Vembar, Olszewski, Subramanyan, Lavi, and Weese]{Ecabert2008}
O.~Ecabert, J.~Peters, H.~Schramm, C.~Lorenz, J.~Von Berg, M.~J. Walker,
  M.~Vembar, M.~E. Olszewski, K.~Subramanyan, G.~Lavi, and J.~Weese.
\newblock Automatic model-based segmentation of the heart in {CT} images.
\newblock \emph{Medical Imaging, {IEEE} Transactions on}, 27\penalty0
  (9):\penalty0 1189--1201, 2008.
\newblock ISSN 0278-0062.
\newblock \doi{10.1109/TMI.2008.918330}.

\bibitem[Glocker et~al.(2012)Glocker, Pauly, Konukoglu, and
  Criminisi]{Glocker2012}
B.~Glocker, O.~Pauly, E.~Konukoglu, and A.~Criminisi.
\newblock Joint {classification}-{regression} {forests} for {spatially}
  {structured} {multi}-object {segmentation}.
\newblock In \emph{Computer {Vision} at {ECCV} 2012}, number 7575 in Lecture
  {Notes} in {Computer} {Science}, pages 870--881. Springer Berlin Heidelberg,
  October 2012.
\newblock ISBN 978-3-642-33764-2, 978-3-642-33765-9.
\newblock \doi{10.1007/978-3-642-33765-9_62}.

\bibitem[Chandra et~al.(2016)Chandra, Dowling, Greer, Martin, Wratten, Pichler,
  Fripp, and Crozier]{Chandra2016a}
S.~S. Chandra, J.~A. Dowling, P.~Greer, J.~Martin, C.~Wratten, P.~Pichler,
  J.~Fripp, and S.~Crozier.
\newblock Fast automated segmentation of multiple objects via spatially
  weighted shape learning.
\newblock \emph{Physics in Medicine and Biology}, 61\penalty0 (22):\penalty0
  8070, 2016.
\newblock \doi{10.1088/0031-9155/61/22/8070}.

\bibitem[Chandra et~al.(2012)Chandra, Dowling, Shen, Raniga, Pluim, Greer,
  Salvado, and Fripp]{Chandra2012c}
S.~S. Chandra, J.~A. Dowling, K.~Shen, P.~Raniga, J.~P.~W. Pluim, P.~B. Greer,
  O.~Salvado, and J.~Fripp.
\newblock {Patient specific prostate segmentation in 3-D magnetic resonance
  images}.
\newblock \emph{Medical Imaging, IEEE Transactions on}, 31\penalty0
  (10):\penalty0 1955 --1964, October 2012.
\newblock ISSN 0278-0062.
\newblock \doi{10.1109/TMI.2012.2211377}.

\bibitem[{Long} et~al.(2015){Long}, {Shelhamer}, and {Darrell}]{Long2015}
J.~{Long}, E.~{Shelhamer}, and T.~{Darrell}.
\newblock Fully convolutional networks for semantic segmentation.
\newblock In \emph{2015 IEEE Conference on Computer Vision and Pattern
  Recognition (CVPR)}, pages 3431--3440, June 2015.
\newblock \doi{10.1109/CVPR.2015.7298965}.

\bibitem[Noh et~al.(2015)Noh, Hong, and Han]{noh2015learning}
H.~Noh, S.~Hong, and B.~Han.
\newblock Learning deconvolution network for semantic segmentation.
\newblock In \emph{Proceedings of the IEEE international conference on computer
  vision}, pages 1520--1528, 2015.

\bibitem[Badrinarayanan et~al.(2017)Badrinarayanan, Kendall, and
  Cipolla]{badrinarayanan2017segnet}
V.~Badrinarayanan, A.~Kendall, and R.~Cipolla.
\newblock Segnet: A deep convolutional encoder-decoder architecture for image
  segmentation.
\newblock \emph{IEEE transactions on pattern analysis and machine
  intelligence}, 39\penalty0 (12):\penalty0 2481--2495, 2017.

\bibitem[Kendall et~al.(2015)Kendall, Badrinarayanan, and
  Cipolla]{kendall2015bayesian}
A.~Kendall, V.~Badrinarayanan, and R.~Cipolla.
\newblock Bayesian segnet: Model uncertainty in deep convolutional
  encoder-decoder architectures for scene understanding.
\newblock \emph{arXiv preprint arXiv:1511.02680}, 2015.

\bibitem[Ronneberger et~al.(2015)Ronneberger, Fischer, and
  Brox]{Ronneberger2015}
O.~Ronneberger, P.~Fischer, and T.~Brox.
\newblock U-net: Convolutional networks for biomedical image segmentation.
\newblock In \emph{International Conference on Medical image computing and
  computer-assisted intervention}, pages 234--241. Springer, 2015.
\newblock ISBN 978-3-319-24573-7.
\newblock \doi{10.1007/978-3-319-24574-4_28}.

\bibitem[Milletari et~al.(2016)Milletari, Navab, and Ahmadi]{milletari2016v}
F.~Milletari, N.~Navab, and S.~Ahmadi.
\newblock {V-net}: Fully convolutional neural networks for volumetric medical
  image segmentation.
\newblock In \emph{2016 Fourth International Conference on 3D Vision (3DV)},
  pages 565--571. IEEE, 2016.

\bibitem[Dong et~al.(2017)Dong, Yang, Liu, Mo, and Guo]{Dong2017}
H.~Dong, G.~Yang, F.e Liu, Y.n Mo, and Y.~Guo.
\newblock Automatic brain tumor detection and segmentation using {U-Net} based
  fully convolutional networks.
\newblock In \emph{Medical Image Understanding and Analysis}, pages 506--517,
  Cham, 2017. Springer International Publishing.
\newblock ISBN 978-3-319-60964-5.
\newblock \doi{10.1007/978-3-319-60964-5_44}.

\bibitem[Norman et~al.(2018)Norman, Pedoia, and Majumdar]{Norman2018}
B.~Norman, V.~Pedoia, and S.~Majumdar.
\newblock Use of {2D} {U-Net} convolutional neural networks for automated
  cartilage and meniscus segmentation of knee {MR} imaging data to determine
  relaxometry and morphometry.
\newblock \emph{Radiology}, 288\penalty0 (1):\penalty0 177--185, 2018.
\newblock ISSN 1527-1315 (Electronic) 0033-8419 (Linking).
\newblock \doi{10.1148/radiol.2018172322}.

\bibitem[{\c{C}}i{\c{c}}ek et~al.(2016){\c{C}}i{\c{c}}ek, Abdulkadir, Lienkamp,
  Brox, and Ronneberger]{cieck2016}
{\"O}.~{\c{C}}i{\c{c}}ek, A.~Abdulkadir, S.~S. Lienkamp, T.~Brox, and
  O.~Ronneberger.
\newblock {3D} {U-Net}: learning dense volumetric segmentation from sparse
  annotation.
\newblock In \emph{International Conference on Medical Image Computing and
  Computer-assisted Intervention}, pages 424--432. Springer, 2016.

\bibitem[Kayalibay et~al.(2017)Kayalibay, Jensen, and van~der
  Smagt]{kayalibay2017cnn}
B.~Kayalibay, G.~Jensen, and P.~van~der Smagt.
\newblock {CNN-based} segmentation of medical imaging data.
\newblock \emph{arXiv preprint arXiv:1701.03056}, 2017.

\bibitem[Isensee et~al.(2017)Isensee, Kickingereder, Wick, Bendszus, and
  Maier-Hein]{isensee2017brain}
F.~Isensee, P.~Kickingereder, W.~Wick, M.~Bendszus, and K.~H. Maier-Hein.
\newblock Brain tumor segmentation and radiomics survival prediction:
  Contribution to the brats 2017 challenge.
\newblock In \emph{International MICCAI Brainlesion Workshop}, pages 287--297.
  Springer, 2017.

\bibitem[He et~al.(2016)He, Zhang, Ren, and Sun]{he2016deep}
K.~He, X.~Zhang, S.~Ren, and J.~Sun.
\newblock Deep residual learning for image recognition.
\newblock In \emph{Proceedings of the IEEE conference on computer vision and
  pattern recognition}, pages 770--778, 2016.

\bibitem[Xue et~al.(2020)Xue, Farhat, Boukrina, Barrett, Binder, Roshan, and
  Graves]{xue2020multi}
Y.~Xue, F.~G. Farhat, O.~Boukrina, A.~M. Barrett, J.~R. Binder, U.~W. Roshan,
  and W.~W. Graves.
\newblock A multi-path 2.5 dimensional convolutional neural network system for
  segmenting stroke lesions in brain {MRI} images.
\newblock \emph{NeuroImage: Clinical}, 25:\penalty0 102118, 2020.

\bibitem[Wang et~al.(2019)Wang, Li, Vercauteren, and
  Ourselin]{wang2019automatic}
G.~Wang, W.~Li, T.~Vercauteren, and S.~Ourselin.
\newblock Automatic brain tumor segmentation based on cascaded convolutional
  neural networks with uncertainty estimation.
\newblock \emph{Frontiers in computational neuroscience}, 13:\penalty0 56,
  2019.

\bibitem[Roth et~al.(2015)Roth, Lu, Liu, Yao, Seff, Cherry, Kim, and
  Summers]{roth2015improving}
H.~R. Roth, L.~Lu, J.~Liu, J.~Yao, A.~Seff, K.~Cherry, L.~Kim, and R.~M.
  Summers.
\newblock Improving computer-aided detection using convolutional neural
  networks and random view aggregation.
\newblock \emph{IEEE transactions on medical imaging}, 35\penalty0
  (5):\penalty0 1170--1181, 2015.

\bibitem[Dolz et~al.(2018)Dolz, Xu, Rony, Yuan, Liu, Granger, Desrosiers,
  Zhang, Ben~Ayed, and Lu]{Dolz2018}
J.~Dolz, X.~Xu, J.~Rony, J.~Yuan, Y.~Liu, E.~Granger, C.~Desrosiers, X.~Zhang,
  I.~Ben~Ayed, and H.~Lu.
\newblock Multiregion segmentation of bladder cancer structures in {MRI} with
  progressive dilated convolutional networks.
\newblock \emph{Med Phys}, 45\penalty0 (12):\penalty0 5482--5493, 2018.
\newblock ISSN 2473-4209 (Electronic) 0094-2405 (Linking).
\newblock \doi{10.1002/mp.13240}.

\bibitem[Dong et~al.(2020)Dong, Luo, Tam, Wang, Wang, Cao, Chen, Zhang, and
  Li]{dong2020deep}
S.~Dong, G.~Luo, C.~Tam, W.~Wang, K.~Wang, S.~Cao, B.~Chen, H.~Zhang, and
  S.~Li.
\newblock Deep atlas network for efficient {3D} left ventricle segmentation on
  echocardiography.
\newblock \emph{Medical image analysis}, 61:\penalty0 101638, 2020.

\bibitem[Sun et~al.(2019)Sun, Fan, Ding, Huang, and Paisley]{Sun2019}
L.~Sun, Z.~Fan, X.~Ding, Y.~Huang, and J.~Paisley.
\newblock Joint {CS-MRI} reconstruction and segmentation with a unified deep
  network.
\newblock In \emph{International Conference on Information Processing in
  Medical Imaging}, pages 492--504. Springer International Publishing, 2019.
\newblock \doi{978-3-030-20351-1}.

\bibitem[Wang et~al.(2020)Wang, Sun, Cheng, Jiang, Deng, Zhao, Liu, Mu, Tan,
  Wang, et~al.]{wang2020deep}
J.~Wang, K.~Sun, T.~Cheng, B.~Jiang, C.~Deng, Y.~Zhao, D.~Liu, Y.~Mu, M.~Tan,
  X.~Wang, et~al.
\newblock Deep high-resolution representation learning for visual recognition.
\newblock \emph{IEEE transactions on pattern analysis and machine
  intelligence}, 2020.

\bibitem[Minaee et~al.(2020)Minaee, Boykov, Porikli, Plaza, Kehtarnavaz, and
  Terzopoulos]{minaee2020image}
S.~Minaee, Y.~Boykov, F.~Porikli, A.~Plaza, N.~Kehtarnavaz, and D.~Terzopoulos.
\newblock Image segmentation using deep learning: A survey.
\newblock \emph{arXiv preprint arXiv:2001.05566}, 2020.

\bibitem[Wang et~al.(2018{\natexlab{a}})Wang, Chen, Yuan, Liu, Huang, Hou, and
  Cottrell]{wang2018understanding}
P.~Wang, P.~Chen, Y.~Yuan, D.~Liu, Z.~Huang, X.~Hou, and G.~Cottrell.
\newblock Understanding convolution for semantic segmentation.
\newblock In \emph{2018 IEEE winter conference on applications of computer
  vision (WACV)}, pages 1451--1460. IEEE, 2018{\natexlab{a}}.

\bibitem[Yang et~al.(2018)Yang, Yu, Zhang, Li, and Yang]{yang2018denseaspp}
M.~Yang, K.~Yu, C.~Zhang, Z.~Li, and K.n Yang.
\newblock Dense {ASPP} for semantic segmentation in street scenes.
\newblock In \emph{Proceedings of the IEEE Conference on Computer Vision and
  Pattern Recognition}, pages 3684--3692, 2018.

\bibitem[Paszke et~al.(2016)Paszke, Chaurasia, Kim, and
  Culurciello]{paszke2016enet}
A.~Paszke, A.~Chaurasia, S.~Kim, and E.~Culurciello.
\newblock Enet: A deep neural network architecture for real-time semantic
  segmentation.
\newblock \emph{arXiv preprint arXiv:1606.02147}, 2016.

\bibitem[Chen et~al.(2014)Chen, Papandreou, Kokkinos, Murphy, and
  Yuille]{chen2014semantic}
L.~Chen, G.~Papandreou, I.~Kokkinos, K.~Murphy, and A.~L. Yuille.
\newblock Semantic image segmentation with deep convolutional nets and fully
  connected {CRFs}.
\newblock \emph{arXiv preprint arXiv:1412.7062}, 2014.

\bibitem[Chen et~al.(2017{\natexlab{a}})Chen, Papandreou, Kokkinos, Murphy, and
  Yuille]{chen2017deeplab}
L.~Chen, G.~Papandreou, I.~Kokkinos, K.~Murphy, and A.~L. Yuille.
\newblock Deeplab: Semantic image segmentation with deep convolutional nets,
  atrous convolution, and fully connected {CRFs}.
\newblock \emph{IEEE transactions on pattern analysis and machine
  intelligence}, 40\penalty0 (4):\penalty0 834--848, 2017{\natexlab{a}}.

\bibitem[Chen et~al.(2017{\natexlab{b}})Chen, Papandreou, Schroff, and
  Adam]{chen2017rethinking}
L.~Chen, G.~Papandreou, F.~Schroff, and H.~Adam.
\newblock Rethinking atrous convolution for semantic image segmentation.
\newblock \emph{arXiv preprint arXiv:1706.05587}, 2017{\natexlab{b}}.

\bibitem[Chen et~al.(2018)Chen, Zhu, Papandreou, Schroff, and
  Adam]{chen2018encoder}
L.~Chen, Y.~Zhu, G.~Papandreou, F.~Schroff, and H.~Adam.
\newblock Encoder-decoder with atrous separable convolution for semantic image
  segmentation.
\newblock In \emph{Proceedings of the European conference on computer vision
  (ECCV)}, pages 801--818, 2018.

\bibitem[Li et~al.(2017)Li, Wang, Fidon, Ourselin, Cardoso, and
  Vercauteren]{li2017compactness}
W.~Li, G.~Wang, L.~Fidon, S.~Ourselin, M.~J. Cardoso, and T.~Vercauteren.
\newblock On the compactness, efficiency, and representation of {3D}
  convolutional networks: brain parcellation as a pretext task.
\newblock In \emph{International conference on information processing in
  medical imaging}, pages 348--360. Springer, 2017.

\bibitem[Yu et~al.(2005)Yu, Wang, and Lai]{yu2005integrated}
L.~Yu, S.~Wang, and K.~K. Lai.
\newblock An integrated data preparation scheme for neural network data
  analysis.
\newblock \emph{IEEE Transactions on Knowledge and Data Engineering},
  18\penalty0 (2):\penalty0 217--230, 2005.

\bibitem[Zhou et~al.(2019)Zhou, Siddiquee, Tajbakhsh, and
  Liang]{zhou2019unet++}
Z.~Zhou, M.~M.~R. Siddiquee, N.~Tajbakhsh, and J.~Liang.
\newblock {Unet++}: Redesigning skip connections to exploit multiscale features
  in image segmentation.
\newblock \emph{IEEE transactions on medical imaging}, 39\penalty0
  (6):\penalty0 1856--1867, 2019.

\bibitem[Holschneider et~al.(1990)Holschneider, Kronland-Martinet, Morlet, and
  Tchamitchian]{holschneider1990CANFRE}
M.~Holschneider, R.~Kronland-Martinet, J.~Morlet, and P.~Tchamitchian.
\newblock A real-time algorithm for signal analysis with the help of the
  wavelet transform.
\newblock In \emph{Wavelets}, pages 286--297. Springer, 1990.

\bibitem[Romera et~al.(2017)Romera, Alvarez, Bergasa, and
  Arroyo]{romera2017efficient}
E.~Romera, J.~Alvarez, L.~M. Bergasa, and R.~Arroyo.
\newblock Efficient convnet for real-time semantic segmentation.
\newblock In \emph{2017 IEEE Intelligent Vehicles Symposium (IV)}, pages
  1789--1794. IEEE, 2017.

\bibitem[Liew et~al.(2016)Liew, Khalil-Hani, and Bakhteri]{liew2016bounded}
S.~Liew, M.~Khalil-Hani, and R.~Bakhteri.
\newblock Bounded activation functions for enhanced training stability of deep
  neural networks on visual pattern recognition problems.
\newblock \emph{Neurocomputing}, 216:\penalty0 718--734, 2016.

\bibitem[Wang et~al.(2018{\natexlab{b}})Wang, Phillips, Sui, Liu, Yang, and
  Cheng]{wang2018classification}
S.~Wang, P.~Phillips, Y.~Sui, B.~Liu, M.~Yang, and H.~Cheng.
\newblock Classification of {Alzheimer’s} disease based on eight-layer
  convolutional neural network with leaky rectified linear unit and max
  pooling.
\newblock \emph{Journal of Medical Systems}, 42\penalty0 (5):\penalty0 85,
  2018{\natexlab{b}}.

\bibitem[{Chen} et~al.(2017){Chen}, {Xu}, and {Koltun}]{Chen2017}
Q.~{Chen}, J.~{Xu}, and V.~{Koltun}.
\newblock Fast image processing with fully-convolutional networks.
\newblock In \emph{2017 IEEE International Conference on Computer Vision
  (ICCV)}, pages 2516--2525, Oct 2017.
\newblock \doi{10.1109/ICCV.2017.273}.

\bibitem[Ulyanov et~al.(2016)Ulyanov, Vedaldi, and
  Lempitsky]{ulyanov2016instance}
D.~Ulyanov, A.~Vedaldi, and V.~Lempitsky.
\newblock Instance normalization: The missing ingredient for fast stylization.
\newblock \emph{arXiv preprint arXiv:1607.08022}, 2016.

\bibitem[Le et~al.(2015)Le, Jaitly, and Hinton]{le2015simple}
Q.~V. Le, N.~Jaitly, and G.~E. Hinton.
\newblock A simple way to initialize recurrent networks of rectified linear
  units.
\newblock \emph{arXiv preprint arXiv:1504.00941}, 2015.

\bibitem[Tran et~al.(2015)Tran, Bourdev, Fergus, Torresani, and
  Paluri]{tran2015learning}
D.~Tran, L.~Bourdev, R.~Fergus, L.~Torresani, and M.~Paluri.
\newblock Learning spatiotemporal features with {3D} convolutional networks.
\newblock In \emph{Proceedings of the IEEE international conference on computer
  vision}, pages 4489--4497, 2015.

\bibitem[Cha(2007)]{cha2007comprehensive}
S.~Cha.
\newblock Comprehensive survey on distance/similarity measures between
  probability density functions.
\newblock \emph{City}, 1\penalty0 (2):\penalty0 1, 2007.

\bibitem[Seim et~al.(2010)Seim, Kainmueller, Lamecker, Bindernagel, Malinowski,
  and Zachow]{seim2010model}
H.~Seim, D.~Kainmueller, H.~Lamecker, M.~Bindernagel, J.~Malinowski, and
  S.~Zachow.
\newblock Model-based auto-segmentation of knee bones and cartilage in {MRI}
  data.
\newblock \emph{Proc. Medical Image Analysis for the Clinic: A Grand Challenge.
  Bejing, China}, pages 215--223, 2010.

\bibitem[Dowling et~al.(2015{\natexlab{b}})Dowling, Sun, Pichler,
  Rivest-H{\'e}nault, Ghose, Richardson, Wratten, Martin, Arm, Best,
  et~al.]{dowling2015automatic}
J.~A Dowling, J.~Sun, P.~Pichler, D.~Rivest-H{\'e}nault, S.~Ghose,
  H.~Richardson, C.~Wratten, J.~Martin, J.~Arm, L.~Best, et~al.
\newblock Automatic substitute {CT} generation and contouring for {MRI}-alone
  external beam radiation therapy from standard {MRI} sequences.
\newblock \emph{Int J. Radiat. Oncol. Biol. Phys}, 93:\penalty0 1144--53,
  2015{\natexlab{b}}.

\bibitem[Tustison et~al.(2010)Tustison, Avants, Cook, Zheng, Egan, Yushkevich,
  and Gee]{tustison2010n4itk}
N.~J. Tustison, B.~B. Avants, P.~A. Cook, Y.~Zheng, A.~Egan, P.~A. Yushkevich,
  and J.~C. Gee.
\newblock {N4ITK}: Improved {N3} bias correction.
\newblock \emph{IEEE transactions on medical imaging}, 29\penalty0
  (6):\penalty0 1310--1320, 2010.

\end{thebibliography}

\end{document}